\documentclass[11pt]{article}

\title{Contagion in financial systems: A Bayesian network approach}

\author{
Carsten Chong\thanks{Center for Mathematical Sciences, Technical University of Munich, Boltzmannstra\ss e 3, 85748 Garching, Germany, e-mail: carsten.chong@tum.de, cklu@tum.de, URL: www.statistics.ma.tum.de}
~and
Claudia Kl\"uppelberg$^\ast$
}
\usepackage{latexsym}
\usepackage{amssymb}
\usepackage{amsmath}
\usepackage{amsthm}
\usepackage{mathrsfs}
\usepackage[titletoc]{appendix}
\usepackage{subcaption}
\usepackage[round]{natbib}
\usepackage{url}
\usepackage{color}
\usepackage{dsfont}
\usepackage{longtable}
\usepackage{lmodern}
\usepackage{tikz}
\usepackage{tkz-graph}
\usepackage[latin1]{inputenc}
\usepackage{enumitem}
\setenumerate{leftmargin=*}
\usepackage{amsopn}
\usepackage{comment}
\usepackage{lmodern}
\usepackage[linktocpage]{hyperref}

\newcommand{\bfi}{\begin{fig}}
\newcommand{\efi}{\end{fig}}
\newcommand{\btab}{\begin{tab}}
\newcommand{\etab}{\end{tab}}
\newcommand{\barr}{\begin{array}}
\newcommand{\earr}{\end{array}}
\newcommand{\beq}{\begin{equation}}
\newcommand{\eeq}{\end{equation}}
\newcommand{\bdis}{\begin{displaymath}}
\newcommand{\edis}{\end{displaymath}\noindent}

\newcommand{\bbp}{\mathbb{P}}

\newcommand{\bone}{\mathds 1}

\newcommand{\lec}{\lesssim}
\newcommand{\halmos}{\quad\hfill $\Box$}

\newcommand{\cals}{{\cal S}}
\newcommand{\calc}{{\cal C}}

\newcommand{\cale}{{\cal E}}

\newcommand{\caln}{{\cal N}}

\newcommand{\cald}{{\cal D}}
\newcommand{\calv}{{\cal V}}

\newcommand{\calo}{{\cal O}}

\newcommand{\si}{{\sigma}}
\newcommand{\Si}{{\Sigma}}
\newcommand{\om}{{\omega}}

\newcommand{\ee}{\mathrm{e}}

\newcommand{\PP}{\mathrm{P}}

\newcommand{\CC}{\mathrm{C}}

\newcommand{\pa}{\mathrm{pa}}

\newcommand{\desc}{\mathrm{de}}
\newcommand{\ndesc}{\mathrm{nd}}
\newcommand{\anc}{\mathrm{an}}
\newcommand{\darrow}{\rightleftharpoons}
\newcommand{\asi}{\mathrm{ASI}}
\newcommand{\rsi}{\mathrm{RSI}}

\makeatletter
\newcommand{\opnorm}{\@ifstar\@opnorms\@opnorm}
\newcommand{\@opnorms}[1]{%
  \left|\mkern-1.5mu\left|\mkern-1.5mu\left|
   #1
  \right|\mkern-1.5mu\right|\mkern-1.5mu\right|
}
\newcommand{\@opnorm}[2][]{%
  \mathopen{#1|\mkern-1.5mu#1|\mkern-1.5mu#1|}
  #2
  \mathclose{#1|\mkern-1.5mu#1|\mkern-1.5mu#1|}
}
\makeatother

\newtheoremstyle{neu}
    {11pt}      
    {11pt}      
    {}                  
    {}          
    {\bfseries} 
    {}          
    {1em}  
    {\textbf{\thmname{#1}\thmnumber{ #2}\thmnote{ (#3)}}}          
    
\newtheoremstyle{proof}
    {11pt}      
    {11pt}      
    {}                  
    {}          
    {\bfseries} 
    {}            
    {1em}          
    {\textbf{\thmname{#1}\thmnote{ #3}.}}           

\newtheorem{Theorem}{Theorem}[section]
\newtheorem{Corollary}[Theorem]{Corollary}
\newtheorem{Lemma}[Theorem]{Lemma}
\newtheorem{Proposition}[Theorem]{Proposition}
\theoremstyle{neu}
\newtheorem{Definition}[Theorem]{Definition}
\newtheorem{Example}[Theorem]{Example}
\newtheorem{Remark}[Theorem]{Remark}
\newtheorem{Assumption}{Assumption}

\theoremstyle{proof}
\newtheorem{Proof}{Proof}

\newcommand{\bthm}{\begin{Theorem}}
\newcommand{\ethm}{\end{Theorem}}

\newcommand{\bcor}{\begin{Corollary}}
\newcommand{\ecor}{\end{Corollary}}

\newcommand{\blem}{\begin{Lemma}}
\newcommand{\elem}{\end{Lemma}}

\newcommand{\bprop}{\begin{Proposition}}
\newcommand{\eprop}{\end{Proposition}}

\newcommand{\bdf}{\begin{Definition}}
\newcommand{\edf}{\end{Definition}}

\newcommand{\bex}{\begin{Example}}
\newcommand{\eex}{\end{Example}}

\newcommand{\brem}{\begin{Remark}}
\newcommand{\erem}{\end{Remark}}

\newcommand{\bass}{\begin{Assumption}}
\newcommand{\eass}{\end{Assumption}}

\newcommand{\bpr}{\begin{Proof}}
\newcommand{\epr}{\end{Proof}}

\newcommand{\benu}{\begin{enumerate}}
\newcommand{\eenu}{\end{enumerate}}

\newcommand{\bit}{\begin{itemize}}
\newcommand{\eit}{\end{itemize}}

\numberwithin{equation}{section}

\allowdisplaybreaks[0]

\begin{document}


\maketitle

\begin{abstract}
We develop a structural default model for interconnected financial institutions in a probabilistic framework. For all possible network structures we characterize the joint default distribution of the system using Bayesian network methodologies. Particular emphasis is given to the treatment and consequences of cyclic financial linkages. We further demonstrate how Bayesian network theory can be applied to detect contagion channels within the financial network, to measure the systemic importance of selected entities on others, and to compute conditional or unconditional probabilities of default for single or multiple institutions. 
\end{abstract}

\vfill

\noindent
\begin{tabbing}
\end{tabbing}

\vspace{1cm}


\vspace{0.5cm}

\newpage

\section{Introduction}\label{Sect1}

The 2007-08 financial crisis has unmistakably revealed on which fragile grounds the global financial system was built at that time. Contrary to the usual perception that diversification reduces risk, it was precisely the interconnectedness of the system that made the spread of shocks across large scales possible. Henceforth, a lot of effort has been put into understanding the origin, the mechanisms and the consequences of financial contagion. It is still a matter of current research to analyze the complex effects of financial interconnection and whether it is a blessing or a curse for the stability of the overall financial system. For example, while the classical papers by \cite{Allen00} and \cite{Freixas00} argue that interconnection strengthens the resilience of financial systems, recent work by \cite{Acemoglu15} and \cite{Elliott14} demonstrate the robust-yet-fragile nature of financial systems: Densely connected systems can prove to be more robust, but also to be more vulnerable to shocks, depending on the actual network structure. Therefore, a profound understanding of interconnectedness and the risk of financial contagion is inevitable for bankers, regulators and other decision makers in order to set up effective detection and prevention mechanisms for possible future crises.

In the literature on balance sheet contagion, one of the most prominent default cascade models is that of \cite{Eisenberg01}. Under mild assumptions on the operative cash flows the financial institutions generate, the authors use a fixed-point argument to establish the existence of a unique clearing payment vector, which can be explicitly computed via a fictitious default algorithm. Extending the model by introducing bankruptcy costs, \cite{Rogers13} show that the clearing vector is no longer uniquely determined in general. Instead, the fictitious default algorithm can be modified to produce the greatest and the least clearing vector. Although \cite{Eisenberg01} also discuss qualitative implications of stochastic operative cash flows, the core analysis of these two papers remains deterministic.

Imposing maximal bankruptcy costs (i.e., creditors lose the total value of a claim upon default of the counterparty), \cite{Gai10} show that the relationship between connectedness and stability in financial systems exhibits a phase transition: Interconnection can both absorb and amplify shocks. Analytical results, however, are only obtained on homogeneous random graphs where exposures are evenly distributed among all counterparties. We refer to \cite{Hurd16} for extensions and comparisons between the Eisenberg-Noe and the Gai-Kapadia model.

A nonmonotonic effect of interconnection on stability is also confirmed by \cite{Acemoglu15}. Drawing on a variant of Eisenberg and Noe's model, they show that for small shocks, highly diversified systems are less susceptible to contagion than sparsely connected networks, but that the opposite is true for big shocks. Methodologically, the authors impose a two-state evolution for the firms and identify the optimal network for the small as well as the big shock regime. Assuming that interconnection arises through asset cross-holdings rather than mutual liabilities, \cite{Elliott14} derive the same qualitative result on various random graph models. 

Another interesting development of the Eisenberg-Noe model is carried out by \cite{Gourieroux12}. The authors first extend the existence and uniqueness of a clearing vector to the situation where interfirm share holdings are included in the model. Additional to these (again purely deterministic) considerations, they further discuss the consequences for the financial system when random shocks occur, giving two quantitative results: First, they compute the conditional distribution of the firm values when it is known which firms default and which not. Second, they provide an enumeration type formula for the probability of default for a given firm in the network. However, they do not reveal how the terms in their expansion (which are the probabilities of joint default configurations) can be computed. Furthermore, it is unclear how to determine which institutions default or not at first place because this in turn depends on the firm values. Finally, let us also mention \cite{Glasserman15} who give quantitative estimates on the probability of default in various contagion models.

In addition to the mainly theoretical contributions mentioned above, a lot of empirical work has been carried out, see e.g. \cite{Elsinger06}, \cite{Cont13}, \cite{Craig14} and \cite{Langfield14} for applications to central bank data from the Austrian, Brazilian, German and UK banking system, respectively. Furthermore, apart from work based on variations of the \cite{Eisenberg01} model, there are also other network-based approaches to systemic risk. We refer to \cite{Castiglionesi15} for the usage of flow network techniques, to \cite{Fouque13}, \cite{Kley15} and \cite{Chong15} for models relying on stochastic differential equations, and to \cite{Amini16} and \cite{Detering15} for contagion analyses on large random graphs. While the analysis in \cite{Castiglionesi15} is again deterministic, the other papers rely on probabilistic approaches that only apply to stylized or limiting networks.

Even though some of the aforementioned works discuss stochastic effects in the context of financial contagion, the full structure of the default probabilities within a given (not necessarily stylized) financial system remains unknown when future returns are random. And this is exactly the problem we want to address in the present paper. For all possible network structures of a financial system subjected to stochastic shocks, we give \emph{a complete characterization of the joint default probability distribution} of the system, in a way that the probability of default for a single or a group of institutions can be computed efficiently. Moreover, our analysis will also allow for identifying possible channels of financial contagion and measuring the systemic importance of institutions. 

Our approach will draw upon methods from the theory of \emph{Bayesian networks}, also known as \emph{directed graphical models}, see e.g.\ \cite{Lauritzen96} and \cite{Koller09}. These models are a priori defined on \emph{directed acyclic graphs (DAGs)}, but it is widely aknowledged that mutual dependencies and directed cycles are omnipresent in financial systems. Therefore, for the purposes of this paper, we have to extend the theory of Bayesian networks to graphs with directed cycles, and we will do so by interpreting a cyclic model as the margin of an acyclic model on a suitably enlarged graph. This methodology for treating cyclic graphical models or, more generally, systems with feedback cycles, is of independent interest and may be useful for other applications  as well.

The remaining article is organized as follows. In Section~\ref{Sect2}, we introduce a probabilistic structural default model for interconnected financial institutions. 
As in \cite{Gai10}, \cite{Rogers13} and \cite{Elliott14}, our model takes bankruptcy costs into account so that the standard balance sheet equations may no longer have a unique solution. Apart from specific examples, a general characterization of uniqueness of solutions seems to be absent in the literature. So this leads to our first main result, Theorem~\ref{uniqueness}, which states that uniqueness holds if and only if the underlying liability network corresponds to a DAG.

If the interfirm liabilities exhibit a DAG structure, we further prove in Section~\ref{Sect3}, Theorem~\ref{PD-DAG}, that the default variables of the institutions form a Bayesian network on this DAG. In other words, if we fix an institution $i$ and assume knowledge about the state of all its debtors (its \emph{parents} in the language of graph theory), its default probability is independent of all other firms that have not lent (directly or indirectly via other firms) to $i$ (the \emph{non-descendants} of $i$).

If there are directed lending cycles, which is most likely the case in practice, the previously identified non-uniqueness problem requires further specification of the sets of defaulting and surviving firms. For the two extremal cases, the \emph{mild} and the \emph{strict default rule}, 
 we recover in Theorems~\ref{Bayes2} and \ref{Bayes} the Bayesian network structure of the default variables on a suitably enlarged liability graph. So for all financial networks---no matter with or without directed cycles---the joint default distribution of the system can be efficiently structured in terms of a Bayesian network. 

In Section~\ref{impl} we discuss the practical consequences and advantages of having a Bayesian network structure. We first reveal how to decide on a graphical basis whether the default of two (or two groups of) firms is stochastically dependent or not. Second, we highlight in a numerical study how established algorithms can be used to compute conditional or unconditional default probabilities, either exactly or approximatively. Finally, we define and analyze two measures of systemic risk  that follow from our Bayesian network analysis. In contrast to many already existing measures, 
ours are probabilistic in nature and follow from a structural default cascade model. 
Section~\ref{Sect5} concludes and points at further research directions.








\section{Default risk model for interconnected firms}\label{Sect2}
We consider $N$ firms labeled $i=1,\ldots,N$ within an economy for a single period.
At time $t=0$, each firm $i$ has some operating assets $X_i$ and some cash $K_i$ that yields a riskless continuous interest at a rate of $r_0$. The same firm also has some external liabilities $F_i$ that are to be redeemed at time $t=T$, the end of the period, with a continuous interest rate of $r^\prime_0$. In addition to that, at $t=0$, each firm $i$ has lent firm $j$ an amount of $L_{ij}$ (possibly $0$) which is to be repaid by $j$ at time $t=T$ with a continuous interest rate of $r_{ij}$. In other words, firm $i$ holds a bond of firm $j$ with face value $L_{ij}$ and a single coupon payment at maturity $T$. Apart from that, no other financial involvement exists between the $N$ firms. 

During the period from $t=0$ to $t=T$ every company $i$ uses its operating assets $X_i$ 
to run a firm-specific business. Following the classical work of \cite{Merton74}, this is assumed to evolve according to a geometric Brownian motion with drift $\mu_i$ and volatility $\si^2_i$. The driving Brownian motions $B_1, \ldots, B_N$ are considered to be independent. We shall comment on possible relaxations of this assumption in Section~\ref{Sect5}. 

In order to simplify the subsequent exposition, the values $K_i$, $F_i$ and $L_{ij}$ are assumed to be nonnegative, the asset drifts $\mu_i$ as well as the interest rates $r_{ij}$, $r_0$ and $r_0^\prime$  may take positive or negative values, and $X_i$, $\si_i$ and $T$ are strictly positive. Moreover, for each $i\in\{1,\ldots,N\}$ we suppose that
	\beq\label{Ki} K_i < \ee^{(r^\prime_0-r_0) T}F_i +\displaystyle\sum_{j=1}^N  \ee^{(r_{ji}-r_0) T} L_{ji}.\eeq 
Firms that have more cash reserves than the bound on the right-hand side of \eqref{Ki} will never experience default at time $t=T$. Even if they do not receive payments from the other institutions and lose all their operating assets, they still have enough money to meet all their obligations. Since we are interested in the default probabilities within the financial network, there is no loss of generality if we exclude such institutions from our analysis.

Let us also point out that we take a ``central bank'' perspective in this paper. All parameters and quantities introduced above are assumed to be known or have been estimated before. We refer to \cite{Anand15}, \cite{Gandy16} and \cite{Upper11} for various methods to estimate interfirm exposures.
\setlength{\unitlength}{0.8cm}
\begin{figure}[!ht]
	\begin{picture}(0,12)
	\put(0,6){
		\begin{tabular}{|p{3.2cm}|p{3.6cm}|}
		\hline
		~&~\\
		Operating assets & Liabilities\\
		~&~- external\\
		$X_i$ & $F_i$\\
		~&~\\
		Loans & - to other firms\\
		$\displaystyle\sum_{j=1}^N L_{ij}$ & $\displaystyle\sum_{j=1}^N L_{ji}$\\
		~&~\\
		Cash&Equity\\
		$K_i$ &$E_i(0)$\\
		~&~\\
		\hline
		~&~\\
		Assets at $t=0$ & Liabilities and equity at $t=0$\\
		~&~\\
		\hline
		\end{tabular}
	}
	
	\put(10.2,6){
		\begin{tabular}{|p{3.2cm}|p{3.6cm}|}
		\hline
		~&~\\
		Operating assets & Liabilities\\
		~&~- external\\
		$X_i(T)$ & $\ee^{r^\prime_0 T}F_i$\\
		~&~\\
		Loans & - to other firms\\
		$\displaystyle\sum_{j=1}^N \ee^{r_{ij} T}L_{ij}\bone_{\cals}(j)$ & $\displaystyle\sum_{j=1}^N \ee^{r_{ji} T}L_{ji}$\\
		~&~\\
		Cash&Equity\\
		$\ee^{r_0 T}K_i$ &$E_i(T)$\\
		~&~\\
		\hline
		~&~\\
		Assets at $t=T$ & Liabilities and equity at $t=T$\\
		~&~\\
		\hline
		\end{tabular}
	}
	\end{picture}
	\caption{Idealized balance sheet of firm $i$ at $t=0$ and $t=T$.}\label{tab1}
\end{figure}

Based on the previously described set-up, the balance sheet of firm $i$ at time $t=0$ has the form depicted on the left-hand side of Figure~\ref{tab1}. In essence, the situation at time $t=0$ is a variant of the classical model by \cite{Eisenberg01}.
At the time of maturity, however, ours differs in two important aspects from that of \cite{Eisenberg01}. 

First, the value of firm $i$'s operating assets at $t=T$, which is given by
\[ X_i(T)=X_i\ee^{(\mu_i-\si^2_i/2)t+\si_i B_i(T)},
\]
is \emph{random}, whereas the operating cash flows in \cite{Eisenberg01} are assumed to be deterministic.

Second, in contrast to a linear or proportional recovery, we impose a \emph{zero recovery rate} in the default scenario; that is, if a firm is not able to meet all of its obligations, none of its creditors receives any payment. As \cite{Gai10} argue, ``this assumption is likely to be realisitic in the midst of a crisis: In the immediate aftermath of a default, the recovery rate and the timing of recovery will be highly uncertain and banks' funders are likely to assume the worst-case scenario'' (p.~2407). We also refer to \cite{Battiston12b} and \cite{Cont13} for a similar reasoning and \cite{Memmel12} for empirical evidence for the frequent occurrence of a recovery rate close to zero.

Standard balance sheet considerations lead us to the following definition (we do not distinguish between the terms bankruptcy, default and insolvency in this paper).
\bdf\label{surv-default} A company \emph{defaults} (\emph{survives}) at time $t=T$ if it belongs to the set
\beq\label{SD}
\cald := \{i\in\{1,\ldots,N\}\colon E_i(T)<0\}\quad \big(\cals := \{i\in\{1,\ldots,N\}\colon E_i(T)\geq 0\}\big),
\eeq
where $E_i(T)$ is the \emph{equity} of firm $i$ at time $T$ given by
\beq
	E_i(T) :=X_i(T) + \sum_{j=1}^N \ee^{r_{ij} T}L_{ij}\bone_{\cals}(j) + \ee^{r_0 T} K_i - \ee^{r^\prime_0 T}F_i- \sum_{j=1}^N \ee^{r_{ji} T}L_{ji}.\label{ET}
\eeq
Moreover, we define
\[ D_i := \bone_{\cald}(i),\quad S_i := 1-D_i = \bone_{\cals}(i),\quad i=1,\ldots,N.  \]
\edf



The resulting balance sheets at time $t=T$ are depicted on the right-hand side of Figure~\ref{tab1}. As a consequence of the stochasticity of the asset values $X_i(T)$, also the equity values $E_i(T)$, the sets $\cald$ and $\cals$ and the variables $D_i$ and $S_i$ are random. 
 The natural question arises whether the knowledge of $X_1(T),\ldots,X_N(T)$ at time $T$ fully determines $\cald$, $\cals$ and $E_i(T)$ through equations \eqref{SD} and \eqref{ET}. In the model of \cite{Eisenberg01} the existence and uniqueness of solutions to \eqref{SD} and \eqref{ET} would immediately follow from the strict positivity of all $X_i(T)$. This condition, however, is no longer sufficient for uniqueness under the zero recovery assumption as the following simple counterexample demonstrates.
\bex\label{twofirms}
Suppose there are two firms in the system owing each other \$10 at maturity with neither cash deposits nor external liabilities. If both firms have realized operating assets amounting to $X_1(T)=X_2(T)=\$5$ at $t=T$, 
there are two different solutions to equations \eqref{SD} and \eqref{ET}:
\benu
	\item $\cald=\{1,2\}$, $\cals=\emptyset$ and $E_1(T)=E_2(T)=-\$5$, and
	\item $\cald=\emptyset$, $\cals=\{1,2\}$ and $E_1(T)=E_2(T)=\$5$.
\eenu
\eex

The emergence of multiple solutions in the previous example highlights the remaining point of indeterminancy in our model: To what extent is clearing between distressed firms, or more precisely, the netting of mutual claims allowed? In the first solution of Example~\ref{twofirms}, netting is forbidden and both firms default because
neither of them is able to initiate the promised payment to the other one. In the second solution, netting is permitted, so both firms abandon their claims and survive with \$5 each. Of course, the degree to which netting of claims occurs depends on many external factors such as the availability of short-term fundings or the degree of counterparty anonymity. However, as we shall see in Section~\ref{Sect3}, it is always possible to find and characterize the solution to \eqref{SD} and \eqref{ET} with the maximal or the minimal number of surviving firms, corresponding to the situation where netting is always permitted or prohibited.

The non-uniqueness issue as encountered in Example~\ref{twofirms} is common to cascade models with bankruptcy costs, cf.\ \cite{Rogers13}, \cite{Elliott14} and \cite{Hurd16}. A general criterion for uniqueness versus non-uniqueness seems to be missing in the literature, and hopeless if it is required to hold $\omega$-wise for all possible outcomes of $X_1(T)(\om),\ldots,X_N(T)(\om)$. 

However, we will show that it is possible to characterize all models where uniqueness holds \emph{almost surely}.
 The most efficient way to do so is to represent the interfirm liabilities by a graphical structure. In this article, we only consider \emph{directed graphs} $G=(\calv,\cale)$ where $\calv$ is some finite \emph{vertex set} and $\cale\subseteq \{(i,j)\colon i,j\in\calv, i\neq j\}$ some \emph{edge set} without self-loops.
 The following graphical representation of the interfirm liabilities plays a crucial role in the default analysis of the firm network described above.
\bdf Abbreviating $[N]:=\{1,\ldots,N\}$, we define the \emph{redemption graph} $G=(\calv,\cale)$ as the graph on the vertex set $\calv:=[N]$ with edges $\cale:=\{(i,j)\in[N]^2\colon L_{ji}>0\}$.
\edf
Thus, there is an edge from $i$ to $j$ precisely when $i$ has borrowed money from $j$ at time $t=0$ and has to repay the latter at time $t=T$. In graphical terms, this is denoted by $i\to_G j$ or $i\in\pa_G(j)$, and we say that $i$ is a \emph{parent} of $j$ in $G$. A whole sequence $i_1\to_G \ldots \to_G i_n$ with $n\geq 2$ is then called a \emph{(directed) path} from $i_1$ to $i_n$, and a \emph{(directed) cycle} if $i_1=i_n$. If $G$ contains no directed cycles, $G$ is called a \emph{directed acyclic graph (DAG)}.

For future reference, let us also introduce some further terminology. A vertex $i$ is called an \emph{ancestor} of $j$, and $j$ a \emph{descendant} of $i$, in short $i\in\anc_G(j)$ and $j\in\desc_G(i)$, if there exists a directed path from $i$ to $j$ in $G$. Nodes in $\ndesc_G(i):=\calv\setminus(\{i\}\cup\desc_G(i))$ are called the \emph{non-descendants} of $i$. Given an additional sequence $(x_i)_{i\in\calv}$, we also write $\pa_G(x_i):=(x_j\colon j\in \pa_G(i))$, and define $\anc_G(x_i)$, $\desc_G(x_i)$ and $\ndesc_G(x_i)$ analogously. If $I\subseteq \calv$, we set $\pa_G(I):=\bigcup_{i\in I} \pa_G(i)$, $\desc_G(I):=\bigcup_{i\in I} \desc_G(i)$, $\ndesc_G(I):=\bigcap_{i\in I} \ndesc_G(i)$ and $x_I:=(x_i\colon i\in I)$.

Returning to the question of uniqueness of solutions to \eqref{SD} and \eqref{ET}, our first main result asserts that the decision can be made by sole inspection of the redemption graph.
\bthm\label{uniqueness} Let $G$ be the redemption graph underlying the considered financial network.
\benu	\item If $G$ is a DAG, there exists a unique solution to \eqref{SD} and \eqref{ET} with probability $1$, so  $\cald$, $\cals$ and $E_1(T),\ldots,E_N(T)$ are uniquely determined for almost all realizations of $X_1(T),\ldots,X_N(T)$.
\item If $G$ is not a DAG but contains a directed cycle, there is a strictly positive probability that the realizations $X_1(T),\ldots,X_N(T)$ are such that \eqref{SD} and \eqref{ET} have multiple solutions.
\eenu
\ethm

\section{Default probabilities for interconnected firms}\label{Sect3}

As shown in Theorem~\ref{uniqueness}, directed cycles in the financial system (as observed in Example~\ref{twofirms}) impose the only hindrance to the uniqueness of solutions to \eqref{SD} and \eqref{ET}. So if the redemption graph $G$ is a DAG, the random variables $D_1,\ldots,D_N$ are well defined, and in this case, their joint distribution has a particularly convenient structure. 
\bdf Given a DAG $G=(\calv,\cale)$, a collection $\{X_i\colon i\in\calv\}$ of random variables taking values in a finite set $E$ is said to form a \emph{Bayesian network} over $G$ if for all $e=(e_i\colon i\in \calv)\in E^{|\calv|}$ we have
\beq\label{BNdef} \bbp[X_\calv = e] = \prod_{i\in\calv} \bbp[X_i=e_i \mid \pa_G(X_i)=\pa_G(e_i)]. \eeq
\edf
For a detailed treatment of Bayesian networks we refer to the monographs \cite{Lauritzen96} and \cite{Koller09}. An equivalent characterization is this (see Theorem~3.27 in \cite{Lauritzen96}): $\{X_i\colon i\in\calv\}$ forms a Bayesian network over $G$ if and only if for every $i\in\calv$, the variable $X_i$ is conditionally independent of $X_{\ndesc_G(i)}$ given $X_{\pa_G(i)}$. Thus, a Bayesian network structure can be understood as a collection of conditional independence statements among variables.
\bthm\label{PD-DAG} If the redemption graph $G$ is a DAG, 
the variables $\{D_1,\ldots,D_N\}$ form a Bayesian network on $G$ with conditional probabilities
\beq\label{condprobDAG} \bbp[D_i=1\mid \pa_G(D_i)] = \Phi(i,\Sigma_i,T),\quad \bbp[D_i=0\mid \pa_G(D_i)] = 1-\Phi(i,\Sigma_i,T),\quad i\in[N], \eeq
where
\begin{align}
\label{Sigmai} \Sigma_i&:=\{j\in\pa_G(i)\colon D_j=0\},\\
\label{Phi}\Phi(i,\Sigma_i,T)&:=\Phi\left(-\frac{\log(Q(i,\Sigma_i))+ (\mu_i-\si^2_i/2)T}{\si_i\sqrt{T}}\right),\\
 Q(i,\Sigma_i)&:=\frac{X_i}{\big(\ee^{r_0^\prime T}F_i  + \sum_{j=1}^N \ee^{r_{ji} T}L_{ji}-\ee^{r_0 T}K_i-\sum_{j\in\Sigma_i} \ee^{r_{ij} T}L_{ij}\big)^+},\label{Qi}\end{align}
and $\Phi$ is the standard normal distribution function, $x^+:=\max(x,0)$ and $a/0:=\infty$ for $a>0$, $\log(\infty):=\infty$, and $\Phi(-\infty):=0$.

In particular, for all $\Sigma\subseteq [N]$ we have that
\beq\label{pd-DAG} \bbp[\cals = \Sigma, ~\cald =[N]\setminus\Sigma] = \prod_{i\in[N]\setminus\Sigma} \Phi(i,\Sigma,T) \prod_{i\in\Sigma} (1-\Phi(i,\Sigma,T)).\eeq
\ethm

Let us give an informal explanation why an acyclic redemption graph leads to a Bayesian network structure of $D_1,\ldots,D_N$. Indeed, every DAG $G$ induces a partial order $\lec_G$ on its vertex set via the relation
\beq\label{lec}  i\lec_G j :\!\!\iff i\in\anc_G(j)\text{ or } i=j. \eeq
Therefore, starting with institutions that are minimal with respect to $\lec_G$ (i.e., institutions that have not lent money to anybody else), we can determine the values of $D_1,\ldots, D_N$ iteratively along the partial order $\lec_G$. The Bayesian network structure follows now from the observation that this iteration procedure is ``Markovian'' with respect to $\lec_G$: For each $i\in[N]$, the variable $D_i$ is independent of $(D_j\colon j\lec_G i)$ given $(D_j\colon j\in \pa_G(i))$, or equivalently, in order to determine the default behavior of $i$, it suffices to know whether its debtors are solvent or not. The formal proof of Theorem~\ref{PD-DAG} can be found in the Appendix. 

When the redemption graph has directed cycles, additional rules to \eqref{SD} and \eqref{ET} have to be set up for $D_1,\ldots,D_N$ to be well defined. These rules may be different for each cycle or vary from firm to firm. 
The following result, however, holds for all \emph{consistent extensions} of \eqref{SD} and \eqref{ET} (i.e., measurable mappings of the values $X_1(T),\ldots, X_N(T)$ to $D_1,\ldots,D_N$ such that \eqref{SD} and \eqref{ET} are valid). Recall that the \emph{subgraph of $G=(\calv,\cale)$ induced by $I\subseteq\calv$} is the graph $G_I:=(I,\cale_I)$ with $\cale_I:=I^2\cap\cale$.
\bprop\label{subDAGs} Formula \eqref{pd-DAG} is valid if both the induced subgraphs $G_\Sigma$ and $G_{[N]\setminus\Sigma}$ contain no directed cycles. In the general case, \eqref{pd-DAG} still holds with a ``$\leq$''-sign instead of equality.
\eprop

Apart from Proposition~\ref{subDAGs}, there is not much more that can be said about a particularly chosen solution to \eqref{SD} and \eqref{ET}. However, further structural results can be obtained for the two ``extreme'' cases, that is, where the set of surviving firms is maximal or minimal, respectively.

\bdf\label{def-mdr}
The \emph{mild default rule} specifies the sets $\cals$ and $\cald$ in the following way.
\bit
\item If the operating assets and cash holdings of some firm $i$ are so low that it cannot meet its obligations at time $t=T$ even if it is paid back by all its debtors, then $i$ defaults and is called a \emph{defaulting firm of the first round}. In this case we write $i\in\cald_1$. If no firm falls into this category, all firms survive and $\cald=\emptyset$ and $\cals=[N]$.
\item  For $n\in\{2,\ldots,N\}$ we call $i\in[N]\setminus\bigcup_{m=1}^{n-1} \cald_m$ a \emph{defaulting firm of the $n$-th round}, denoted $i\in\cald_n$, if at time $t=T$ firm $i$ is insolvent given it receives all contractual payments except for those from their debtors in $\bigcup_{m=1}^{n-1} \cald_m$.
\item As soon as $\cald_n=\emptyset$, we set $\cald_{n+1}=\ldots=\cald_N=\emptyset$, $\cald=\bigcup_{m=1}^{n-1} \cald_m$ and $\cals=[N]\setminus\cald$.
\eit
\edf
\bdf\label{def-sdr}
Under the \emph{strict default rule} the sets $\cals$ and $\cald$ are determined as follows.
\bit
\item Every firm $i$ whose operating assets and cash holdings at time $t=T$ are large enough to repay its creditors (regardless whether it receives payments from the other firms) survives. In this case, we call $i$ a \emph{surviving firm of the first round} and write $i\in\cals_1$. If no firm meets this criterion, all firms default: $\cals=\emptyset$ and $\cald=[N]$.
\item For every $n\in\{2,\ldots,N\}$ we call $i\in[N]\setminus\bigcup_{m=1}^{n-1} \cals_m$ a \emph{surviving firm of the $n$-th round}, denoted $i\in\cals_n$, if at time $t=T$ its operating assets, its cash holdings and the payments it receives from firms in $\bigcup_{m=1}^{n-1} \cals_m$ are high enough to redeem its debts. 
\item As soon as $\cals_n=\emptyset$, we set $\cals_{n+1}=\ldots=\cals_N=\emptyset$, $\cals=\bigcup_{m=1}^{n-1} \cals_m$ and $\cald=[N]\setminus\cals$.
\eit 
\edf
Both the mild and the strict default rules are variants of the fictitious default algorithm of \cite{Eisenberg01} and \cite{Rogers13}, the loss cascades described in \cite{Cont13}, the cascade hierarchies of \cite{Elliott14}, or the clearing algorithms of \cite{Kusnetsov16}. 

\bprop\label{prop-sdr} We state some immediate consequences of Definitions~\ref{def-mdr} and \ref{def-sdr}.
\benu
\item The sets $\cals$ and $\cald$ specified by the mild or the strict default rule are consistent extensions of \eqref{SD} and \eqref{ET}. 
\item For every consistent extension of \eqref{SD} and \eqref{ET}, the survival set contains the one prescribed by the strict default rule and is included in the one given by the mild default rule. Similarly, the default set under a consistent extension is always a superset of the default set under the mild default rule and a subset of the default set under the strict default rule.
\item Under the mild (strict) default rule, if $\Sigma\subseteq[N]$ and $G_{[N]\setminus\Sigma}$ ($G_\Sigma$) has no directed cycles, then formula \eqref{pd-DAG} is valid.
\eenu
\eprop 

Apart from the special cases discussed in the previous results, the presence of cycles in $G$ prevents us from a concise description of the joint probability distribution of $D_1,\ldots,D_N$ using Bayesian network methods. However, under the mild or the strict default rule, the Bayesian network structure can be recovered if we ``blow up'' the redemption graph in a suitable way. The idea here is to explicitly incorporate the cascade mechanism behind the mild or the strict default rule into the graphical structure. Instead of a single variable $D_i$ that determines whether firm $i$ defaults or not, we keep track of a whole tuple of variables $D_{in}$, $n\in[N]$, that depend on whether $i$ defaults in the $n$-th round. In fact, to reduce the number of new variables as far as possible, it suffices to consider this separately for the \emph{strongly connected components} of the redemption graph $G$. These are the equivalence classes induced by the equivalence relation
\[i \sim_G j :\!\!\iff i\lec_G j\text{ and } j\lec_G i, \quad i,j\in[N]; \] 
or in economic terms, these are the maximal subsets of firms where any two of them have a direct or an indirect lender--borrower relationship via intermediate firms. 

\bdf\label{augG}
Let $C_1,\ldots,C_m$ be the strongly connected components in $G$ and $N_i$ the size of the strongly connected component firm $i$ belongs to.
The \emph{acyclic augmentation of $G$} is the graph $\bar G:=(\bar \calv,\bar \cale)$ with the vertex set
	\beq\label{barv} \bar \calv:=\{(i,n)\colon i\in[N], n\in[N_i]\}\eeq 
	and the edge set
	\begin{align} \bar \cale:=&\left\{\big((j,N_j),(i,n)\big)\colon j\in\pa_G(i)\cap\ndesc_G(i), n=1,\ldots,N_i\right\}\nonumber\\
	&\cup \bigcup_{i=1}^N \left\{\big((i,n),(i,n+1)\big)\colon N_i>1, n=1,\ldots,N_i-1\right\}\nonumber\\
	&\cup\bigcup_{k=1}^m \left\{\big((j,n),(i,n+1)\big)\colon i,j \in C_k, j\in\pa_G(i), n=1,\ldots, N_i-1\right\} \nonumber\\
	&\cup\bigcup_{k=1}^m \left\{\big((j,n),(i,n+2)\big)\colon i,j \in C_k, j\in\pa_G(i), N_i>2, n=1,\ldots, N_i-2\right\}. \label{Es}\end{align}
\edf
In words, we first create $N_i$ vertex copies in $\bar G$ for each firm $i\in[N]$, numbered from $(i,1)$ to $(i,N_i)$. Then the first type of edges in \eqref{Es} connects two directly linked firms $i$ and $j$ in $G$ that belong to \emph{different} strongly connected components: Namely, if $j$ has to repay $i$ at time $T$, the \emph{last} copy of $j$ is connected to \emph{each} copy of $i$. The second type of linkages connects consecutive copies of a fixed firm. The third (resp. fourth) type of edges joins consecutive (resp. next-but-one) copies of two firms that are directly linked to each other in $G$ and belong to the \emph{same} strongly connected component. 

\begin{figure}[ht]
	\subcaptionbox{$G$}[.4\linewidth]{
		\vspace{3\baselineskip}
		\begin{tikzpicture}
		\SetGraphUnit{2} 		
		\Vertex{1} \EA(1){4} \SO(1){2} \SO(4){3} 
		\tikzset{EdgeStyle/.style = {->}}
		\Edges(1,2,3,4,1)
		\SetGraphUnit{1.5} 
		\WE(1){5} \NO(1){6} 
		\Edges(1,5)
		\Edges(6,1)
		\SetGraphUnit{1} 
		\SOWE(2){7} \NOEA(4){9} \SOEA(3){8}
		\Edges(3,8)
		\Edges(2,7)
		\Edges(4,9)
		\end{tikzpicture}}
	\subcaptionbox{$\bar G$}[.6\linewidth]{
		\begin{tikzpicture}
		\SetGraphUnit{2} 		
		\Vertex[x=0,y=0]{11} \EA(11){12} \EA(12){13} \EA(13){14} 
		\SetGraphUnit{1} 
		\Vertex[x=0,y=-1.5]{21} 
		\SetGraphUnit{2}
		\EA(21){22} \EA(22){23} \EA(23){24} 
		\SetGraphUnit{1} 
		\Vertex[x=0,y=-3]{31} 
		\SetGraphUnit{2}
		\EA(31){32} \EA(32){33} \EA(33){34} 
		\SetGraphUnit{1} 
		\Vertex[x=0,y=-4.5]{41} 
		\SetGraphUnit{2}
		\EA(41){42} \EA(42){43} \EA(43){44} 
		\SetGraphUnit{2}
		\Vertex[x=3,y=2]{61}
		\SetGraphUnit{2}
		\EA(24){71}
		\NO(71){51}
		\EA(34){81}
		\SO(81){91}
		
		\tikzset{EdgeStyle/.style = {->}}
		\Edges(11,12,13,14,51)
		\Edges(21,22,23,24,71)
		\Edges(31,32,33,34,81)
		\Edges(41,42,43,44,91)
		
		\Edges(61,11)
		\Edges(61,12)
		\Edges(61,13)
		\Edges(61,14)
		
		\Edges(11,22) \Edges(12,23) \Edges(13,24)
		\Edges(21,32) \Edges(22,33) \Edges(23,34)
		\Edges(31,42) \Edges(32,43) \Edges(33,44)
		\Edges(41,12) \Edges(42,13) \Edges(43,14)
		
		\Edges(11, 23) \Edges(12, 24)
		\Edges(21, 33) \Edges(22, 34)
		\Edges(31, 43) \Edges(32, 44)
		\Edges(41, 13) \Edges(42, 14)
		\end{tikzpicture}}
	\caption{A cyclic redemption graph $G$ and its acyclic augmentation $\bar G$.}\label{Gs}
\end{figure}
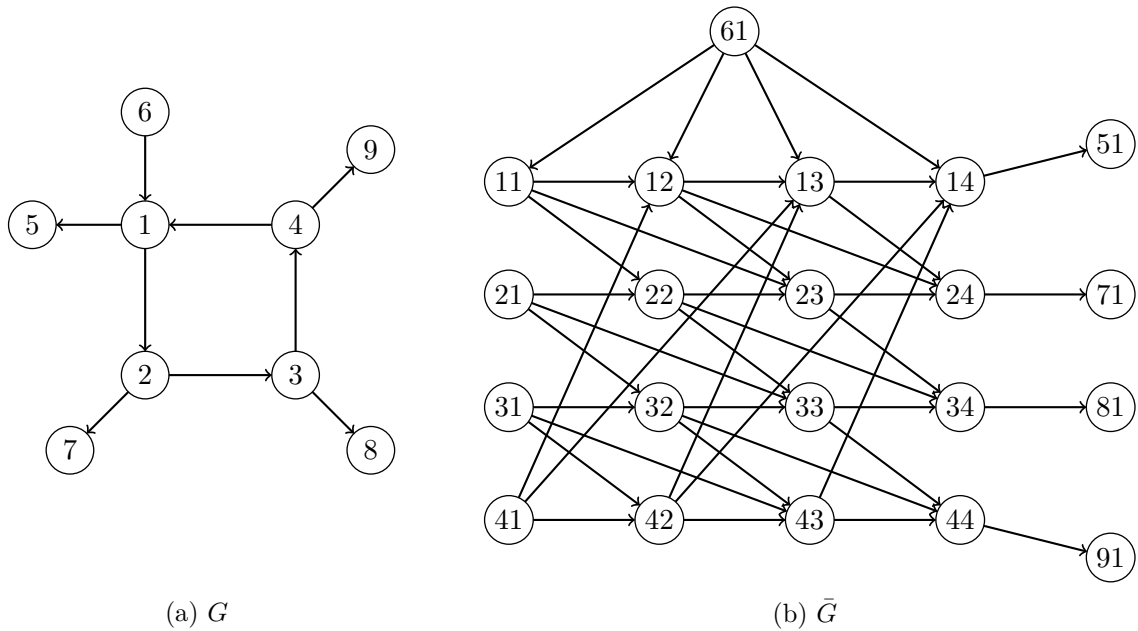

Figure~\ref{Gs} shows an example of a cyclic graph and its acyclic augmentation. As we can see, although $G$ has a directed cycle (the vertices $\{1,2,3,4\}$ form a strongly connected component), its augmentation $\bar G$ has no cycles. Of course, this holds in general.
\blem\label{barG-DAG}
The acyclic augmentation $\bar G$ of the redemption graph $G$ is a DAG.
\elem		

Next we associate random variables to the vertices of $\bar G$, in such a way that they contain our variables of interest $D_1,\ldots, D_N$ as subsets and form a Bayesian network on $\bar G$. We first assume the mild default rule; for the strict default rule, all definitions and results are analogous and will be provided afterwards.

\bdf\label{Din} Let $N_i$ be as in Definition~\ref{augG}. For $i\in [N]$ and $n\in[N_i]$, we define $D_{i0}:=0$ and then inductively $D_{in}=1$ if
\begin{align} &X_i(T) + \ee^{r_0 T} K_i+\sum_{j\in\pa_G(i)\cap\ndesc_G(i)} \ee^{r_{ij} T}L_{ij}\bone_{\{D_{jN_j}=0\}}+\sum_{j\in \pa_G(i)\cap \desc_G(i)} \ee^{r_{ij} T}L_{ij}\bone_{\{D_{j,n-1}=0\}}\nonumber\\
&\quad- \ee^{r_0^\prime T}F_i - \sum_{j=1}^N \ee^{r_{ji} T}L_{ji} < 0, \label{Sin2}\end{align}
and $D_{in}=0$ otherwise. 
\edf
To paraphrase, we first order the strongly connected components of $G$ in such a way that firms in a given component have no borrowings from firms in previous components. Then, starting with those components that have no borrowings from any other component, we assign the value $1$ to $D_{in}$ exactly when firm $i$ defaults assuming it only receives payments from surviving firms of previous strongly connected components and from firms $j$ in the same component that have not defaulted up to the $(n-1)$-st step, that is, where $D_{j,n-1}=0$.

\bthm\label{Bayes2} Under the mild default rule the following statements are valid.
\benu
\item For all $i\in[N]$ and $n\in[N_i]$ we have $D_{i,n-1}\leq D_{in}$ and $D_i = D_{iN_i}$.
\item The random variables $\{D_{in}\colon i\in[N],~n\in[N_i]\}$ form a Bayesian network on $\bar G$ with conditional probabilities given by
\begin{align} \bbp[D_{i1}=1\mid \pa_{\bar G}(D_{i1})] &= \Phi(i,\Si^\mathrm{m}_{i1},T),\nonumber\\
\bbp[D_{in}=1\mid \pa_{\bar G}(D_{in})] &= \begin{cases} 1&\text{if } D_{i,n-1}=1,\\ \frac{\Phi(i,\Si^\mathrm{m}_{in},T)-\Phi(i,\Si^\mathrm{m}_{i,n-1},T)}{1-\Phi(i,\Si^\mathrm{m}_{i,n-1},T)}
&\text{if } D_{i,n-1}=0,\end{cases}\quad n=2,\ldots,N_i,\label{condprob-mdr} \end{align}
where
\[\Si^\mathrm{m}_{in}:=\{j\in\pa_G(i)\cap\ndesc_G(i)\colon D_{jN_j}=0\}\cup \{j\in\pa_G(i)\cap\desc_G(i)\colon D_{j,n-1}=0\},\quad n\in[N_i].\]
\eenu
\ethm

Although the variable $D_{in}$ does not explicitly depend on $D_{i,n-2}$ in Definition~\ref{Din}, we have to form the next-but-one edges in $\bar G$ (i.e., the last of the four types of edges in \eqref{Es}). The reason behind is that on the event $D_{i,n-1}=0$, the values $D_{j,n-2}$ of parents $j$ belonging to the same strongly connected component as $i$ contain important information about $X_i(T)$. For example, if $D_{j,n-2}=1$ for many such parents (so $i$ survives in round $n-1$ even though many of its parents have defaulted in round $n-2$), the value $X_i(T)$ has to exceed a higher threshold compared to a situation where $D_{j,n-2}=0$ for many $j$. Furthermore, the differences $D_{j,n-1}-D_{j,n-2}$ influence the likelihood whether $D_{i,n-1}=D_{in}=0$ or $0=D_{i,n-1} <  D_{in} = 1$. In fact, if $D_{j,n-1}-D_{j,n-2}=0$ for all parents $j$, then necessarily the first alternative occurs. By contrast, the likelihood for the second alternative increases with the frequency that $D_{j,n-1}-D_{j,n-2}=1$ occurs. These considerations are reflected in formula \eqref{Bayes2} via the dependence on $\Si^\mathrm{m}_{i,n-1}$, which involves the variables $(D_{j,n-2}\colon j\in\pa_G(i)\cap\desc_G(i))$.

Next we present analogous statements for the strict default rule, which are simpler to formulate in terms of survival variables.

\bdf
Under the strict default rule we define for $i\in[N]$ and $n\in[N_i]$ the variables $S_{i0}:=0$, and $S_{in}:=1$ if
\begin{align} &X_i(T) +\ee^{r_0 T} K_i+ \sum_{j\in\pa_G(i)\cap\ndesc_G(i)} \ee^{r_{ij} T}L_{ij}\bone_{\{S_{jN_j}=1\}}+\sum_{j\in \pa_G(i)\cap \desc_G(i)} \ee^{r_{ij} T}L_{ij}\bone_{\{S_{j,n-1}=1\}} \nonumber\\
&\quad- \ee^{r_0^\prime T}F_i -\sum_{j=1}^N \ee^{r_{ji} T} L_{ji} \geq0, \label{Sin}\end{align}
and $S_{in}:=0$ otherwise.
\edf

\bthm\label{Bayes} Under the strict default rule the following statements are valid.
\benu
	\item For all $i\in[N]$ and $n\in[N_i]$ we have $S_{i,n-1}\leq S_{in}$ and $S_i = S_{iN_i}$.
	\item The random variables $\{S_{in}\colon i\in[N],~n\in[N_i]\}$ form a Bayesian network on $\bar G$ with conditional probabilities given by
	\begin{align} \bbp[S_{i1}=0\mid \pa_{\bar G}(S_{i1})] &= \Phi(i,\Si^\mathrm{s}_{i1},T),\nonumber\\
	\bbp[S_{in}=0\mid \pa_{\bar G}(S_{in})] &= \begin{cases} 0&\text{if } S_{i,n-1}=1,\\ \frac{\Phi(i,\Si^\mathrm{s}_{in},T)}{\Phi(i,\Si^\mathrm{s}_{i,n-1},T)}&\text{if } S_{i,n-1}=0,\end{cases}\quad n=2,\ldots,N_i,\label{condprob-sdr} \end{align}
	where
	\[\Si^\mathrm{s}_{in}:=\{j\in\pa_G(i)\cap\ndesc_G(i)\colon S_{jN_j}=1\}\cup \{j\in\pa_G(i)\cap\desc_G(i)\colon S_{j,n-1}=1\},\quad n\in[N_i].\]
\eenu
\ethm

The quintessence of this section is that the default variables $D_1,\ldots,D_N$ associated to the financial institutions form a Bayesian network in the case of an acyclic redemption graph, and that they can be embedded as margins into a larger random vector with a Bayesian network structure in the cyclic case. But of course, these results are obtained under the hypotheses on the financial system made in Section~\ref{Sect2}. The question how modifications of these assumptions affect the Bayesian network structure will be addressed in Section~\ref{Sect5}.

\section{Implications of the Bayesian network structure}\label{impl}

We have proved in Theorems~\ref{PD-DAG}, \ref{Bayes2} and \ref{Bayes} that the default and survival random variables associated to firms with mutual liabilities form a Bayesian network over the redemption graph or a suitable augmentation thereof. In this section, we elaborate the benefits of this Bayesian network structure for systemic risk analysis. 

\subsection{Independence detection}

It is important for both the involved institutions and regulatory authorities to determine whether the default of different firms depends on each other or not, and how this relationship is affected by the state of other firms. The underlying Bayesian network structure permits us to reduce this question to a graph-theoretic analysis. More precisely, given the evidence of $D_k$, $k\in K$, for some subset $K\subseteq [N]$, one can determine by sole inspection of $\bar G$, the augmentation of the redemption graph $G$ from Definition~\ref{augG}, whether $\{D_i\colon i\in I\}$ and $\{D_j\colon j\in J\}$ for two further subsets $I,J\subseteq [N]$ depend on each other or not. 

The key concept here is called d-separation, see p.~48 in \cite{Lauritzen96}:
\bdf Let $G=(\calv,\cale)$ be a graph.
\benu
\item We call $i_1 \darrow_G \ldots \darrow_G i_n$ a \emph{chain} between $i_1$ and $i_n$ if for every $j\in [n-1]$ we have $i_j \to_G i_{j+1}$ or $i_{j+1}\to_G i_j$ (or both).
\item For three pairwise disjoint subsets $V_0,V_1,V_2$ of $\calv$, the sets $V_1$ and $V_2$ are called \emph{d-separated in $G$ given $V_0$} (or simply \emph{d-separated} if $V_0=\emptyset$) if every chain $i_1 \darrow_G \ldots \darrow_G i_n$ in $G$ from some $i_1\in V_1$ to some $i_n\in V_2$ is \emph{blocked by $V_0$}, that is, for some $j=2,\ldots, n-1$ we have
\bit
	\item $i_{j-1} \to_G i_j \to_G i_{j+1}$ and $i_j\in V_0$ (\emph{type I structure}), or
	\item $i_{j-1} \leftarrow_G i_j \leftarrow_G i_{j+1}$ and $i_j\in V_0$ (\emph{type II structure}), or
	\item $i_{j-1} \leftarrow_G i_j \to_G i_{j+1}$ and $i_j\in V_0$ (\emph{type III structure}), or
	\item $i_{j-1} \to_G i_j \leftarrow_G i_{j+1}$ and $i_j\notin V_0$ and $\desc_G(i_j)\cap V_0=\emptyset$ (\emph{type IV structure}).
\eit
\eenu
\edf
For example, if $G$ is a DAG, then for every $i\in\calv$ the sets $\{i\}$ and $\ndesc_G(i)$ are d-separated in $G$ given $\pa_G(i)$. If random variables $\{X_i\colon i\in\calv\}$ form a Bayesian network on $G$, the fundamental result is the following, see Corollary~3.23 together with Proposition~3.25 in \cite{Lauritzen96}: 
If $V_1$ and $V_2$ are d-separated in $G$ given $V_0$, then the two collections of random variables $\{X_i\colon i\in V_1\}$ and $\{X_i\colon i\in V_2\}$ are conditionally independent given $\{X_i\colon i\in V_0\}$. Numerical algorithms for checking d-separation are well established and efficient, see Section~3.3 of \cite{Koller09}. 

Applied to the acyclic augmentation $\bar G=(\bar\calv,\bar\cale)$ of the redemption graph, the d-separation criterion provides a graphical tool for detecting independence between different groups of firms regarding their default behavior. In fact, d-separation is ``almost'' necessary for independence (see Theorem~7 of \cite{Meek95}): If $V_0$, $V_1$ and $V_2$ are subsets of $\bar\calv$, and $V_1$ and $V_2$ are not d-separated given $V_0$, then for almost all volatility parameters $\si_i$ the random variables $\{D_i\colon i\in V_1\}$ and $\{D_i\colon i\in V_2\}$ are stochastically dependent given $\{D_i\colon i\in V_0\}$ if none of the variables $D_i$, $i\in V_1\cup V_2$, becomes deterministic given $\{D_i\colon i\in V_0\}$. The next proposition relates d-separation in $G$ to d-separation in $\bar G$.
\bprop\label{dsepeq} Let $G$ be the redemption graph and $\bar G$ its acyclic augmentation. Then the following assertions hold for all pairwise disjoints subsets $V_1$, $V_2$ and $V_3$ of $[N]$.
\benu
	\item $V_1$ and $V_2$ are d-separated in $G$ if and only if $\{(i,N_i)\colon i \in V_1\}$ and $\{(i,N_i)\colon i\in V_2\}$ are d-separated in $\bar G$.
	\item If $V_1$ and $V_2$ are d-separated in $G$ given $V_0$, then $\{(i,n)\colon i\in V_1,~n\in[N_i]\}$ and $\{(i,n)\colon i\in V_2,~n\in[N_i]\}$ are d-separated in $\bar G$ given $\{(i,n)\colon i\in V_0,~n\in[N_i]\}$.
\eenu
\eprop

\subsection{Measures of systemic impact}\label{msi}

In the systemic risk literature several measures of systemic importance for interconnected firms have been proposed. While some approaches like in \cite{Acemoglu15}, \cite{Battiston12b} and \cite{Diebold14} are purely based on the underlying network structure and parameters, 
others explicitly take the probabilistic nature of contagion into account. Examples for the latter category include the contagion index of \cite{Cont13}, the CoVaR measure of \cite{Adrian15}, the SES and SRISK measures of \cite{Acharya10} and \cite{Brownlees15}, and the general systemic risk measures of \cite{Biagini15} and \cite{Feinstein16}. 
 Based on the structural cascade model analyzed in Sections~\ref{Sect2} and \ref{Sect3}, we contribute two further probabilisitic measures of systemic risk that are flexible (one can measure the impact of a single firm or a group of firms, unconditionally or conditionally on different stress scenarios), feasible (they can be computed by the methods described in the next subsection) and robust (they depend continuously on the underlying model parameters). 

In the following, if not otherwise stated, any consistent extension of \eqref{SD} and \eqref{ET} can be considered, in particular both the mild and the strict default rule.
\bdf\label{SysImp} For two disjoint subsets $I$ and $J$ of $[N]$, the \emph{absolute} and the \emph{relative systemic impact (ASI and RSI)} of $I$ on $J$ are defined as (we use $1$ to denote a vector with all entries equal to one)
\begin{align}\label{AbsSysImp}
\asi(I,J) &:= \max_{J_0\subseteq J,~ e \in \{0,1\}^{|J_0|}} \big(\bbp[ D_{J_0} = e \mid D_I = 1] - \bbp[D_{J_0} = e]\big)\\
\text{and}\quad\rsi(I,J)&:=\max_{e \in \{0,1\}^{|J|}} \log_2 \frac{\bbp[ D_J = e \mid D_I = 1]}{\bbp[D_J = e]}, \label{RelSysImp}
\end{align}
respectively. In \eqref{RelSysImp}, we set $\log_2 0=-\infty$, and $0/0=1$ in case the denominator, and, as a consequence, also the numerator is zero. 
\edf

The ASI of $I$ on $J$ equals the \emph{total variation distance} between the distribution of $D_J$ and its conditional distribution given $D_I=1$. Moreover, the RSI of $I$ on $J$ is the \emph{R\'enyi divergence of order $+\infty$} between the same pair of distributions. Of course, instead of taking the maximum in \eqref{AbsSysImp} and \eqref{RelSysImp}, also $L^p$-type measures can be investigated. For example, one can define $L^1$-analogues of the ASI and the RSI measures based on \emph{mutual information} and the \emph{Kullback-Leibler divergence}. We do not go into details at this point but only refer to \cite{vanErven14} for more information on distances between probability measures. 

Turning to the literature on Bayesian networks, we found, to our surprise, only little research where measures of importance for vertices in a network are investigated. Most notably, in Chapter~4 of \cite{Pinto86} (cf. \cite{Pearl88}, Chapter~6.4), the author proposes quantifying the relevance of a vertex $Y$ for another vertex $X$ by a function depending on the distribution of $X$ and its conditional distribution given $Y$. They give a version of the ASI measure as an example. Furthermore, 
\cite{Chan05} employ a symmetric version of the RSI measure in the context of sensitivity analysis for Bayesian networks.

Next, we collect some basic properties of the ASI and RSI measures.
\bprop\label{SysImpProp} The following assertions hold for disjoint sets $I,J\subseteq[N]$.
\benu
\item $\asi(I,J)\in[0,1)$ and $\rsi(I,J) \in [0,\infty)$.
\item If $J_1\subseteq J$, then $\asi(I,J_1)\leq \asi(I,J)$ and $\rsi(I,J_1)\leq \rsi(I,J)$.
\item An explicit formula for \eqref{AbsSysImp} is
\begin{align}  \asi(I,J)&=\sum_{e\in\{0,1\}^{|J|}} \big(\bbp[D_J=e \mid D_I=1] - \bbp[D_J=e]\big)^+ \nonumber\\
&=\frac{1}{2} \sum_{e\in\{0,1\}^{|J|}} \big|\bbp[D_J=e \mid D_I=1] - \bbp[D_J=e]\big|.
\label{ASI-expl}\end{align}
\item In contrast to \eqref{AbsSysImp}, there is no need to consider subsets of $J$ in \eqref{RelSysImp} because
\beq\label{RSI-var} \rsi(I,J)=\max_{J_0\subseteq J,~ e \in \{0,1\}^{|J_0|}} \log_2 \frac{\bbp[ D_{J_0} = e \mid D_I = 1]}{\bbp[D_{J_0} = e]}. \eeq
\item Under the strict default rule, if $I=I_1\cup I_2$ and $I_2$ and $J$ are d-separated in $G$ given $I_1$, then $\asi(I,J)=\asi(I_1,J)$ and $\rsi(I,J)=\rsi(I_1,J)$.
\eenu
\eprop

\begin{figure}[ht]
	\begin{center}
		\begin{tikzpicture}
		\Vertex[x=0,y=0]{1} 
		\Vertex[x=0,y=-2]{2}
		\Vertex[x=3,y=0]{4}
		\Vertex[x=3,y=-2]{3}
		\Vertex[x=6,y=-1]{5}
		\tikzset{EdgeStyle/.style={->,above,sloped,midway}}
		\Edge[label=\$20,labelstyle={above}](1)(4)
		\Edge[label=\$10,labelstyle={below}](2)(3)
		\Edge[label=\$10,labelstyle={above}](1)(3)
		\Edge[label=\$15,labelstyle={above}](4)(5)
		\Edge[label=\$15,labelstyle={below}](3)(5)		
		\end{tikzpicture}
		\vspace{-1\baselineskip}
	\end{center}
	\caption{Redemption graph for the firm network of Example~\ref{samepar}.}\label{5firms}
\end{figure}
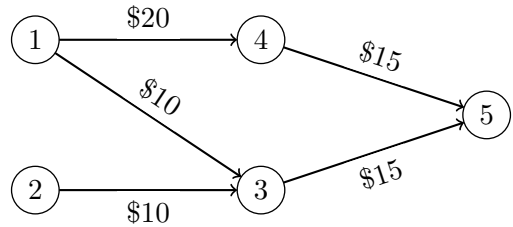

One would also expect $\asi(I_1,J)\leq \asi(I,J)$ and $\rsi(I_1,J)\leq \rsi(I,J)$ to hold for $I_1\subseteq I$ since the knowledge of default of a larger set of firms ought to be worse, or at least not better, for the survival of another firm in the network. However, a simple counterexample demonstrates that this intuition is wrong.
\bex\label{samepar}
We consider a network of $N=5$ institutions with redemption graph as in Figure~\ref{5firms} and zero interest rates. 
Further suppose that $F_3=F_4=K_3=K_4=\$0$, that is, both firm $3$ and firm $4$ have neither cash nor external liabilities at time $t=T$. Moreover, we make the following assumptions on the firms' operative performance.
\bit
\item $\mu_1=\mu_2$ are very high and $\si_1=\si_2$ are strictly positive but small. 
\item $\mu_3$ and $\mu_4$ are strongly negative, and $\sigma_3$ and $\sigma_4$ are small in comparison.
\item $X_1=X_2$ such that $X_1(T)$ and $X_2(T)$ are independent with the same distribution.
\eit

As a consequence of the first hypothesis, the probabilities of default for firms $1$ and $2$ are very small but strictly positive. Moreover, by the second assumption, without repayment from firms $1$ and $2$, firms $3$ and $4$ have almost no chance to survive and pay back firm $5$. However, as soon as firm $1$ survives, also firm $4$ survives, and as soon as both firm $1$ and firm $2$ are solvent, also firm $3$ is so. Hence we obtain: 
\bit
\item $\bbp[D_4=1] \approx 0$ because firm $1$ has a very high chance of survival such that also firm $4$ is most likely to survive.
\item $\bbp[D_4=1 \mid D_3=1] \approx 1/2$ because conditional on the default of firm $3$, the event of exactly one failure among $1$ and $2$ is almost one (the probability of both firms surviving is zero because otherwise $3$ would not have defaulted, and the probability of a joint default of $1$ and $2$ is still comparatively small). Since only the failure of firm $1$ triggers the default of $4$, and $X_1(T)$ and $X_2(T)$ are identically distributed, the conditional default probability of $4$ given the default of $3$ is around $1/2$.
\item $\bbp[D_4=1 \mid D_2=D_3=1] \approx 0$ because the failure of $2$ already explains the failure of $3$, so the latter gives us almost no additional information about $1$ (and therefore $4$).
\eit
As a result, $\asi(\{2,3\},\{4\})\leq \asi(\{3\},\{4\})$ and $\rsi(\{2,3\},\{4\})\leq \rsi(\{3\},\{4\})$.
\eex

\subsection{Computation of default probabilities} 
In addition to finding the dependence relations among interconnected networks, another central objective of risk management is to determine the default probabilities of firms or the impact of a firm's default on other firms, measured, for example, in terms of the ASI and RSI of the previous subsection. Known as the \emph{probabilistic inference problem} in the artificial intelligence community, this topic has been extensively studied in the literature of Bayesian networks, where various exact or approximative algorithms have been developed for this purpose (see \cite{Koller09} for an in-depth treatment on this subject). In order to illustrate the results and concepts of the previous sections, we demonstrate in a numerical example how these inference algorithms can be applied to our setting. All numerical computations in the following are made using the \emph{junction tree algorithm} 
 within the open-source Bayes Net Toolbox for Matlab \citep{Murphy01}. 

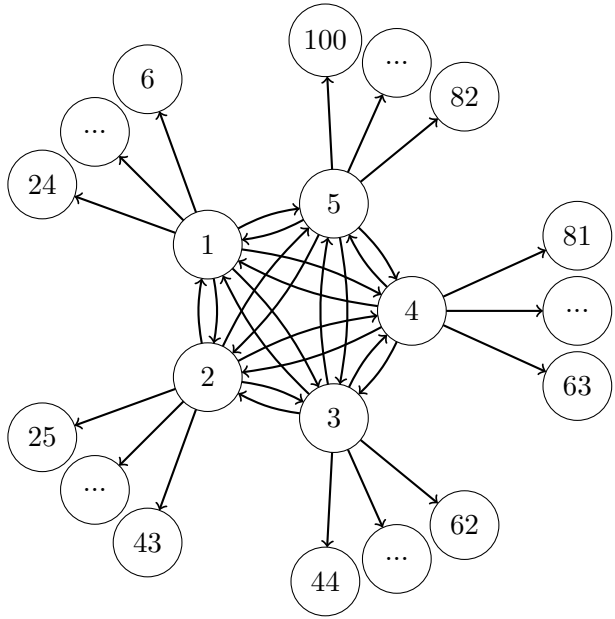
\begin{figure}[t]
	\begin{center}
		\begin{tikzpicture}[rotate=0]
		\SetGraphUnit{1.5} 	
		
		\tikzset{VertexStyle/.style = {shape=circle, minimum size = 26pt, draw}}
		
		\Vertices{circle}{4,5,1,2,3}	
		\tikzset{EdgeStyle/.style = {->,bend left=10}}
		\Edges(1,2,3,4,5,1,4,2,5,3,1,3,5,2,4,1,5,4,3,2,1)
		\NOWE[L=...](1){M1}	
		\SOWE[L=...](2){M2}
		
		\Vertex[x=1.3,y=-3.3,L=...]{M3}
		\Vertex[x=0.35,y=-3.6]{44}
		\Vertex[x=2.2,y=-2.85]{62}
		
		\Vertex[x=1.3,y=3.3,L=...]{M5}
		\Vertex[x=0.35,y=3.6]{100}	
		\Vertex[x=2.2,y=2.85]{82}
		
		\SetGraphUnit{2.2}
		\EA[L=...](4){M4}
		
		\SetGraphUnit{0.7} 	
		\SOWE(M1){24}
		\NOEA(M1){6}
		\NOWE(M2){25}
		\SOEA(M2){43}
		
		\SetGraphUnit{1} 	
		\SO(M4){63}
		\NO(M4){81}
		
		\tikzset{EdgeStyle/.style = {->}}
		\Edges(1,6)
		\Edges(1,M1)
		\Edges(1,24)
		\Edges(2,25)
		\Edges(2,M2)
		\Edges(2,43)
		\Edges(3,44)
		\Edges(3,M3)
		\Edges(3,62)
		\Edges(4,81)
		\Edges(4,M4)
		\Edges(4,63)
		\Edges(5,82)
		\Edges(5,M5)
		\Edges(5,100)
		
		\end{tikzpicture}
	\end{center}
	\caption{A core--periphery structured redemption graph.}\label{cp-ex}
\end{figure}

\begin{figure}[t]
	\includegraphics[width=\linewidth]{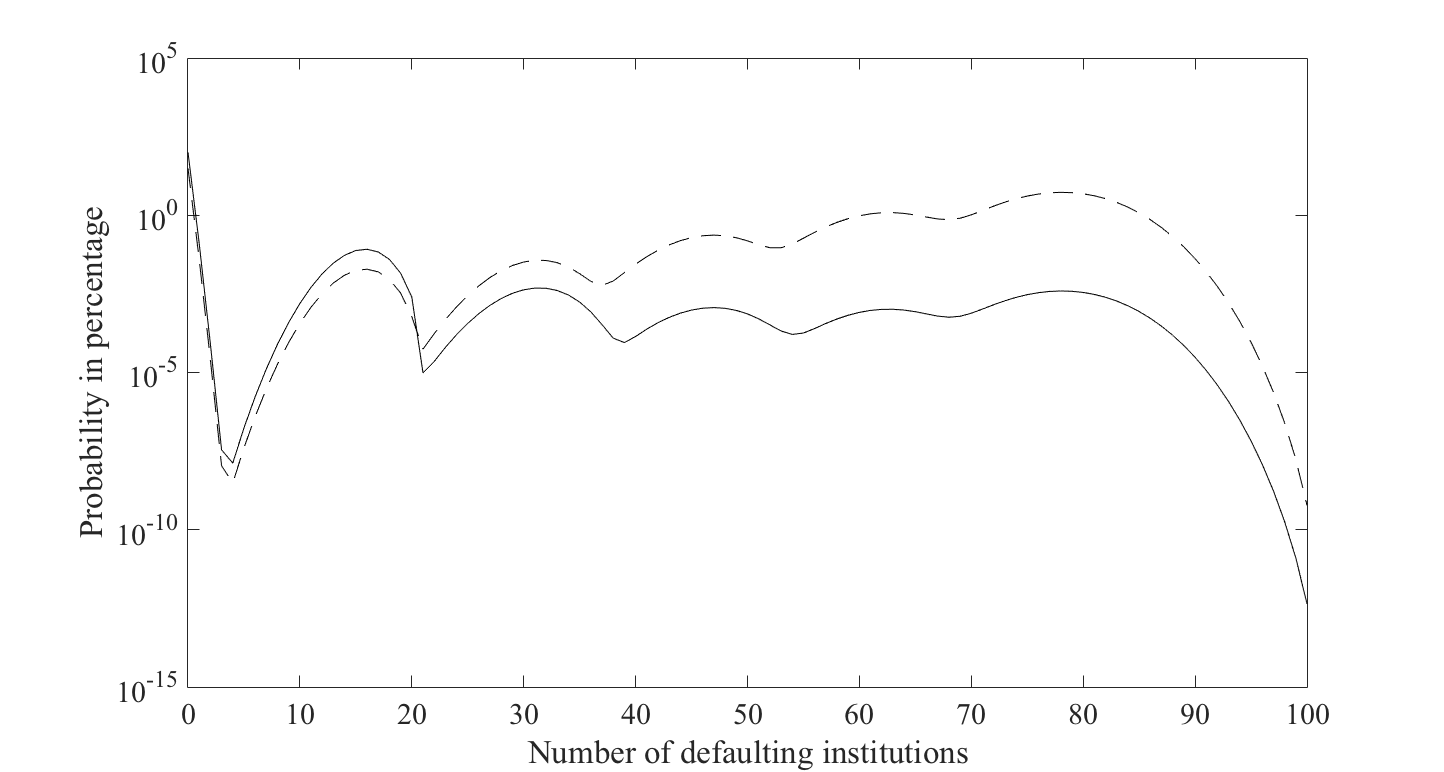}
	\caption{Default probabilities under the mild default rule (solid line) and the strict default rule (dashed line).} \label{numdef1}
\end{figure}

We consider a financial network with $N=100$ banks in total where the redemption graph is depicted in Figure~\ref{cp-ex}. 
The network has a core--periphery structure in the sense that the first five banks form a complete network where all possible lending relationships are present, and each of these core banks has further borrowed money from 19 different periphery banks. We refer to the papers \cite{Elsinger06}, \cite{Cont13}, \cite{Craig14} and \cite{Langfield14} for empirical evidence of such structures in real banking networks. The time horizon in our example is $T=1$ year, all interest rates are zero, and we assume statistical equality among all core banks and among all periphery banks:
\begin{align*}
\mu_\CC&=0.1, &\si_\CC&=0.2, &X_\CC&= 2000, &F_\CC&=500, &L_{\CC \CC} &= 1600/(N-1)=400, \nonumber\\
\mu_\PP&=0.05, &\si_\PP& = 0.1, &X_\PP &= 80, &F_\PP &=90, &L_{\PP \CC} &= 35 ~(\text{if } \CC \to \PP).
\end{align*}
The subscript $\CC$ ($\PP$) here represents an arbitrary core (periphery) banks.
Figure~\ref{numdef1} displays the probability distribution of the number of defaults in the resulting financial systems under the mild and the strict default rule.
While the shape of the default distribution is the same for both the mild and the strict default rule, the actual probabilities are very different. No default occurs under the mild rule with a probability of $99.39\%$, but this probability decreases to $31.15\%$ when the strict rule applies. This discrepancy is due to the fact that a core bank only has a $11.13\%$ chance to survive without the repayment of the other core banks, but a $99.90\%$ chance if it receives full repayment. As a consequence, in the considered example, the possibility of netting mutual claims is decisive whether the contagion cascade becomes unleashed or not.

From Figure~\ref{numdef1} we can observe several prominent qualitative features of the default distribution. First of all, the distribution is multimodal, corresponding to the number of core banks in default. Indeed, conditional on the default of a core bank, each periphery bank with lendings to this core bank defaults (independently) with a probability of $76.66\%$. Thus, whenever a core bank defaults, it triggers the default of $19\times 76.66\%=14.57$ periphery banks on average, which explains why for both the mild and the strict rule the default distribution has modes at distances of $15$ or $16$ (one core bank plus $14$ or $15$ periphery banks). 

\begin{figure}[htb]
	\includegraphics[width=\linewidth]{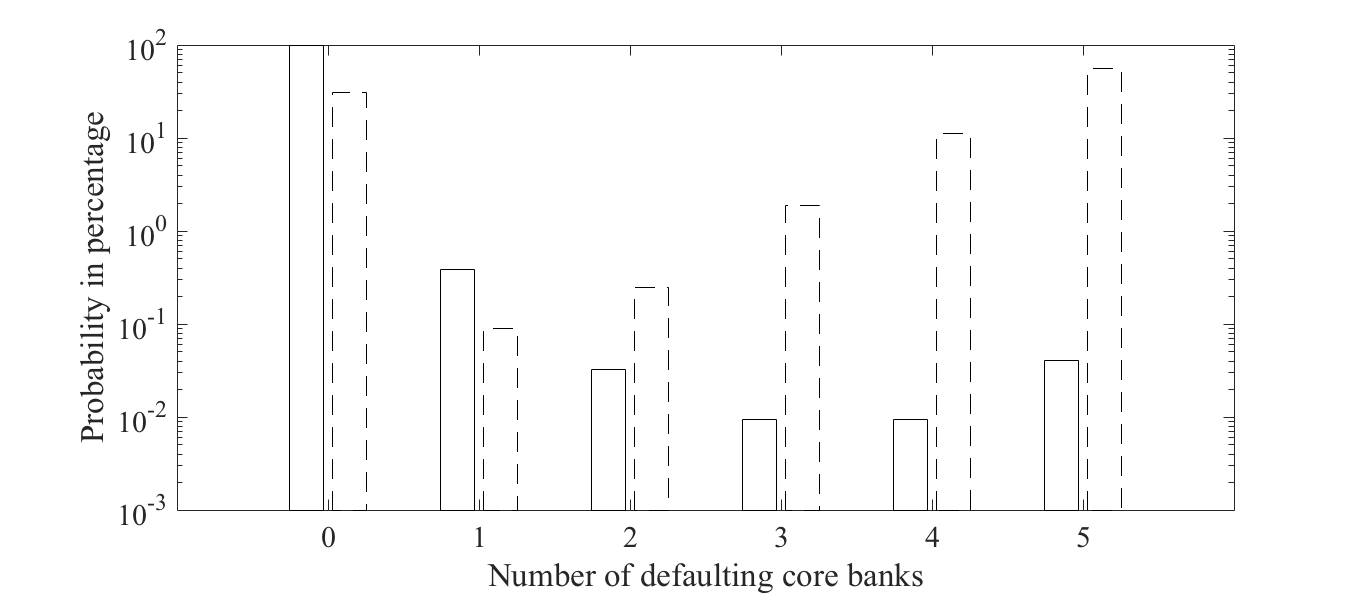}
	\caption{Distribution of the number of defaulting core banks under the mild rule (solid bars) and the strict rule (dashed bars).}\label{coredef}
\end{figure}

Furthermore, we observe that also the probabilities at the modes are not monotone. This is caused by contagious effects among the core banks. When enough core banks default, it is rather likely that even more core banks default: The probability function of the number of defaulting core banks is not monotone as Figure~\ref{coredef} reveals.

Last but not least, the probabilities beyond the last mode with 78 defaults fall into the large deviation regime and decrease exponentially fast.

The measures of systemic impact introduced in the previous subsection can also be calculated in our example. For simplicity, we only report figures for the mild default rule; the computations under the strict default rule are completely analogous. To this end, let $\CC$ and $\CC'$ be two arbitrary core banks and $\PP$ and $\PP'$ be two periphery banks that are creditors of $\CC$ and $\CC'$, respectively. For the absolute systemic impact of $\CC$ we obtain
\begin{align*}
&\asi(\{\CC\}, \{\CC'\}) = \bbp[D_{\CC'} = 1 \mid D_\CC = 1] - \bbp[D_{\CC'} = 1] = 36.21\% - 0.14\% = 36.07\%,\\
&\asi(\{\CC\}, \{\PP\}) = \bbp[D_\PP = 1 \mid D_\CC = 1] - \bbp[D_\PP = 1] = 76.66\% - 0.11\% = 76.55\%,\\
&\asi(\{\CC\}, \{\PP'\}) = \bbp[D_{\PP'} = 1 \mid D_\CC = 1] - \bbp[D_{\PP'} = 1] = 27.76\% - 0.11\% = 27.65\%,
\end{align*}
which translates to the relative systemic impact of $\CC$ as
\[
\rsi(\{\CC\}, \{\CC'\}) = 7.97,\quad \rsi(\{\CC\}, \{\PP\})  = 9.42,\quad \rsi(\{\CC\}, \{\PP'\}) = 7.96. 
\]
Concering the ASI and RSI of periphery banks, we fix another periphery bank $\PP''$ that is, like $\PP$, also a creditor of $\CC$. Then we obtain
\begin{align*}
&\asi(\{\PP\}, \{\CC\}) = \bbp[D_{\CC} = 1 \mid D_\PP = 1] - \bbp[D_{\CC} = 1] = 98.79\% - 0.14\% = 98.65\%,\\
&\asi(\{\PP\}, \{\CC'\}) = \bbp[D_{\CC'} = 1 \mid D_\PP = 1] - \bbp[D_{\CC'} = 1] = 35.78\% - 0.14\% = 35.63\%,\\
&\asi(\{\PP\}, \{\PP'\}) = \bbp[D_{\PP'} = 1 \mid D_\PP = 1] - \bbp[D_{\PP'} = 1] = 27.43\% - 0.11\% = 27.32\%,\\
&\asi(\{\PP\}, \{\PP''\}) = \bbp[D_{\PP''} = 1 \mid D_\PP = 1] - \bbp[D_{\PP''} = 1] = 75.74\% - 0.11\% = 75.63\%,
\end{align*}
and
\begin{align*} &\rsi(\{\PP\}, \{\CC\}) = 9.42,\quad \rsi(\{\PP\}, \{\CC'\})  = 7.96,\quad \rsi(\{\PP\}, \{\PP'\}) = 7.94,\\
 &\rsi(\{\PP\}, \{\PP''\}) = 9.40.  \end{align*}

As both the ASI and RSI reveal, even banks that are not direct counterparties may be systemically important for each other. For instance, if a core bank defaults, the default probability of a periphery bank increases by $27.65\%$ (or by a factor of $2^{7.96} \approx 248$) even if it has no lendings to that defaulted core bank. Another observation is that also periphery banks have a high systemic impact although they have no borrowings from other banks in the system. The reason behind is simply that periphery banks that are repaid by their core bank counterparty survive with a probability larger than $99.99\%$. So if a periphery bank defaults, this is very likely caused by the default of its core bank counterparty.

\section{Conclusion and outlook}\label{Sect5}
In this paper we analyzed a financial contagion model in which the joint default probability distribution of the system can be characterized in terms of a Bayesian network on a graph determined by the interfirm liabilities. We further explained how this graphical structure can be employed to detect systemic dependencies within the network, to define measures of systemic importance for institutions and to compute their conditional or unconditional default probabilities. Since the involved methodologies apply to all possible networks, this work provides a useful device to analyze and monitor the systemic risk in financial systems.

Naturally, we did impose various conditions on the financial network, so let us comment on possible generalizations. First of all, regarding probabilistic assumptions, the log-normal hypothesis on the evolution of the operating firm assets is merely for simplicity. The Bayesian network structure of the default variables is preserved if we assume another distribution for the asset returns. Only the conditional probabilities stated in Theorems~\ref{PD-DAG}, \ref{Bayes2} and \ref{Bayes} have to be adapted. 

The situation is different if we allow for dependent asset returns of the financial institutions. Of course, if one is only interested in computing default probabilities using Monte Carlo techniques, this is still feasible as long as the joint asset return distribution can be simulated from. However, the Bayesian network structure of the default variables $D_1,\ldots,D_N$ will be lost in general because observations of non-descendant institutions (with respect to the redemption graph) may reveal additional information about the driving noise. But let us suppose that each institution $i$ is driven by a finite number of common market noises $W_1,\ldots,W_K$ plus its own idiosyncratic noise $B_i$, assuming that the $B_i$'s are still independent and also independent of the $W_j$'s. Then conditional on the $W_j$'s, the variables $D_1, \ldots, D_N$ will have a Bayesian network structure. Hence, our approach permits default risk analysis under various market situations, which is useful for stress testing, for instance.

Other generalizations may concern the assumptions on the financial system itself and the way institutions are interconnected. In our model, all mutual exposures are loans with two possible values at the end of the period (depending on whether the counterparty fails or not). It would be more realistic to include other types of exposures: For example, exposures with continuously varying values like asset cross-holdings (cf. \cite{Gourieroux12} and \cite{Elliott14}), exposures with different seniorities, exposures with a non-zero recovery rate in the case of counterparty default, or exposures with different maturity dates. These considerations clearly go beyond the scope of the current paper and are left to future research.


\begin{appendix}
\section*{Appendix: Proofs}
\bpr[of Theorem~\ref{uniqueness}] (1)\quad If $G$ is a DAG, the relation $\lec_G$ defined in \eqref{lec} induces a partial order on $\{1,\ldots,N\}$ with a non-empty set $\calv_1$ of minimal nodes with respect to $\lec_G$, which precisely contains those firms that have not lent money to any other firm. For such a firm $i\in\calv_1$, the equation for $E_i(T)$ in \eqref{ET} is explicit: It does not depend on any other $j\neq i$ because $L_{ij}=0$. Next, we consider the minimal nodes in $\calv\setminus\calv_1$ with respect to $\lec_{G_{\calv\setminus\calv_1}}$, that is, those firms that have only lent money to firms in $\calv_1$. Their values $E_i(T)$ are given by an expression involving $E_j(T)$ only for $j\in\calv_1$. Iterating this at most $N$ times, we obtain an explicit formula for all $E_i(T)$ in terms of $X_1(T),\ldots,X_N(T)$ and the model parameters. 

(2)\quad Now we suppose that $G$ has a directed cycle, say, involving the firms $1,\ldots, N_0$ in that order. 
By assumption \eqref{Ki}, for each firm $i\in \{N_0+1, \ldots, N\}$ there exists a range $[x_i^{(1)},x_i^{(2)})$ with $0=x_i^{(1)}<x_i^{(2)}$ such that $i$ necessarily defaults whenever $X_i(T)$ falls into this interval. For the $N_0$ firms within the cycle we first define $L_0$ to be the smallest amount a firm in $\{1,\ldots,N_0\}$ has to repay to its creditor within the cycle. An inspection of \eqref{ET} reveals that there exists a rectangle $\prod_{i=1}^{N_0} [x_i^{(1)},x_i^{(2)})$ with $x_i^{(1)}<x_i^{(2)}$ and the following property: When $(X_1(T),\ldots,X_{N_0}(T))$ lies in that rectangle, then each firm in $[N_0]$ is able to repay the external liabilities and all its creditors except the one in the cycle, but the remaining asset value is larger than the debt owed to this cycle creditor minus $L_0$. As a result, if $(X_1(T),\ldots,X_N(T)) \in \prod_{i=1}^N [x_i^{(1)}, x_i^{(2)})$, which is an event with strictly positive probability, then one solution would be that all firms default, and another would be that the firms in the cycle survive while all other firms default.\halmos
\epr

\bpr[of Theorem~\ref{PD-DAG}]
This theorem is a special case of Theorem~\ref{Bayes}.
\halmos
\epr

\bpr[of Proposition~\ref{subDAGs}] That \eqref{pd-DAG} holds with ``$\leq$'' is immediate: If $\cals=\Sigma$, equation \eqref{ET} becomes an explicit expression in terms of $X_i(T)$ and the model parameters for each $i\in [N]$. The independence of $X_1(T),\ldots,X_N(T)$ together with their log-normal distributions yields \eqref{pd-DAG} with ``$\leq$''. In order to prove equality when both $G_\Sigma$ and $G_{[N]\setminus\Sigma}$ are DAGs, we consider the situation where $X_1(T),\ldots,X_N(T)$ satisfy \eqref{SD} and \eqref{ET} with $\cals=\Sigma$. We have to show that all other choices $\cals=\Sigma^\prime$ with $\Sigma^\prime\neq \Sigma$ would violate \eqref{SD} and \eqref{ET}. Since $G_\Sigma$ is a DAG, the relation $\lec_{G_{\Sigma}}$ induces a partial order on $\Sigma$. The minimal nodes in $\Sigma$ with respect to that partial order, denoted by $\Sigma_1$, correspond to firms that survive without receiving any payment from other firms in the network at time $T$. So they must belong to $\Sigma^\prime$ whenever $\cals=\Sigma^\prime$ is supposed to satisfy \eqref{SD} and \eqref{ET}. Next, we consider the minimal nodes in $\Sigma\setminus \Sigma_1$. Again, they must belong to $\Sigma^\prime$ because they survive regardless of the solvency of all other firms except for the minimal nodes, which, however, are already known to survive. Inductively, this proves that $\Sigma\subseteq\Sigma^\prime$ for every potential $\Sigma^\prime$. Similarly, because $G_{[N]\setminus\Sigma}$ is a DAG, the maximal elements in $[N]\setminus\Sigma$ default regardless whether the other firms in $[N]\setminus\Sigma$ default or not, so they cannot belong to $\Sigma^\prime$. An analogous induction argument yields $[N]\setminus\Sigma\subseteq [N]\setminus\Sigma^\prime$, so $\Sigma=\Sigma^\prime$.\halmos
\epr

\bpr[of Proposition~\ref{prop-sdr}] We assume the strict default rule; for the mild default rule, the proof is completely analogous. 

(1)\quad It is obvious that the strict default rule maps $X_1(T),\ldots,X_N(T)$ in a measurable way to $S_1,\ldots, S_N$. So we are left to prove that for $i\in[N]$ the term in \eqref{ET} is nonnegative if and only if $i\in\cals$. By Definition~\ref{def-sdr}, the expression \eqref{ET} for $i\in\cals_n$ is already nonnegative when we take the indicator $\bone_{\cals_{n-1}}$ (for $n=1$ we set $\cals_0:=\emptyset$), which is always less or equal to the original indicator $\bone_\cals$. Since $\cals=\cals_1\cup\ldots\cup\cals_N$, this proves that \eqref{ET} is nonnegative for all $i\in\cals$. By contrast, for all $i\in\cald$ equation \eqref{ET} must take a strictly negative value because otherwise $i$ belongs to $\cals_1\cup\ldots\cup\cals_{N_0+1}$ according to the strict default rule where $N_0$ is the smallest number in $[N]$ such that $\cals=\cals_1\cup\ldots\cup\cals_{N_0}$, which is obviously a contradiction.


(2)\quad Denote by $\cals_1,\ldots,\cals_N$ the surviving firms of the $n$-th round under the strict default rule. As in (1), a simple induction argument shows that $\cals_n\subseteq \cals$ holds for all $n\in[N]$ whenever $\cals$ and $\cald$ satisfy \eqref{SD} and \eqref{ET}. 

(3)\quad We have already shown in Proposition~\ref{subDAGs} that \eqref{pd-DAG} always holds with ``$\leq$''. For the reverse inequality, suppose that $X_1(T),\ldots,X_N(T)$ satisfy \eqref{SD} and \eqref{ET} with $\Sigma$ instead of $\cals$. Under the strict default rule, the proof of the same proposition reveals that, because of the DAG structure of $G_\Sigma$, we necessarily have $\Sigma\subseteq\cals$, so part (2) of the current proposition completes the proof. 
\halmos\epr

\bpr[of Lemma~\ref{barG-DAG}] Suppose that $\bar G$ does have a cycle, say, of the form 
\[(i_1,n_1)\to_{\bar G} \ldots\to_{\bar G}(i_k,n_k)\to_{\bar G} (i_1,n_1).\]
Then at least one of the edges must belong to the first of the four collections of edges in \eqref{Es} because otherwise we would have $n_1<n_2<\ldots<n_k<n_1$, which is absurd. But if the cycle has an edge of the form $((i_l,N_{i_l}),(i_{l+1},n_{l+1}))$ with $i_l\in\pa_G(i_{l+1})\cap\ndesc_G(i_{l+1})$, there cannot exist a path in $\bar G$ that connects $(i_{l+1},n_{l+1})$ to $(i_l,N_{i_l})$ because, by \eqref{Es}, this would give rise to a path in $G$ from $i_{l+1}$ to $i_l$ and contradict the fact that $i_l\in\ndesc_G(i_{l+1})$. 
However, this in turn produces a contradiction to the assumption that $(i_1,n_1),\ldots,(i_k,n_k)$ form a cycle in $\bar G$. \halmos
\epr

\bpr[of Theorem~\ref{Bayes2}]
(1)\quad 
The first statement follows immediately from the definition in \eqref{Sin2}. 
For the second statement we number the strongly connected components of $G$ in such a way that no vertex in $C_k$ has parents in $C_{k+1},\ldots,C_m$. By induction on $k$ we may assume that $D_i=D_{iN_i}$ holds under the mild default rule for all $i\in C_1\cup\ldots\cup C_{k-1}$. We must prove that $D_i=D_{iN_i}$ is true for $i\in C_k$. We first show that $D_i=1$ holds as soon as $D_{in}=1$ for some $n\in[N_i]$. To this end, we proceed with another induction on $n$, and assume that we already know that $D_{j,n-1}=1$ implies $D_j=1$ for all $j\in C_k$. Then firms $i\in C_k$ with $D_{in}=1$ cannot repay their debts if they only receive payments from $j\in\pa_G(i)\setminus C_k$ with $D_{jN_j}=0$ and $j\in\pa_G(i)\cap C_k$ with $D_{j,n-1}=0$. Indeed, all other firms default under the mild default rule: Firms $j\in\pa_G(i)\setminus C_k$ with $D_{jN_j}=1$ by the first, and firms $j\in\pa_G(i)\cap C_k$ with $D_{j,n-1}=1$ by the second induction hypothesis. Thus, also $i$ must default according to the mild default rule. For the converse direction that $D_i=1$ implies $D_{iN_i}=1$ for all $i\in C_k$, we recall that $\cald=\bigcup_{n=1}^N \cald_n$, where $\cald_n$ contains the defaulting firms of the $n$-th round. By induction on $n\in[N]$, let us assume that those $i\in C_k$ that belong to $\cald_1\cup\ldots\cup\cald_{n-1}$ satisfy $D_{iN_i}=1$. Then every $i\in\cald_n$ 
has a negative equity value at time $T$ if it does not receive money back from $j\in\pa_G(i)\cap(\cald_1\cup\ldots\cup\cald_{n-1})$. If such a firm $j$ belongs to $\pa_G(i)\setminus C_k$, we have $D_{jN_j}=1$ by the first induction hypothesis, while if it belongs to $\pa_G(i)\cap C_k$, we have $D_{jN_j}=1$ by the second hypothesis, and consequently $D_{j,N_j-1}=1$ (because the event that $D_{j,N_j-1}=0$ and $D_{jN_j}=1$ happens only if $j$ defaults in a later round than all firms in $C_k$, which, however, is not possible because $j\in\cald_1\cup\ldots\cup\cald_{n-1}$ and $i\in\cald_n$). Therefore, we conclude that $D_{iN_i}=1$, which ends the proof of $D_i=D_{i N_i}$ for $i\in C_k$.

(2)\quad 
If $\{e_{in}\colon i\in[N],~n\in[N_i]\}$ is a given sequence taking values $0$ or $1$, then, because of (1), the probability that $D_{in}=e_{in}$ for all $i$ and $n$
can only be larger than zero if $e_{in}\leq e_{i,n+1}$ for all $i\in[N]$ with $N_i>1$ and $n\in[N_i-1]$. In this case, we define for $i\in[N]$ the number
\[ n_i:=\begin{cases} \min\{n\in[N_i]\colon e_{in}=1\} &\text{if } e_{iN_i}=1\\ N_i+1 &\text{if } e_{iN_i}=0 \end{cases}\] 
and observe that $D_{in}=e_{in}$ holds for all $i\in[N]$ and $n\in[N_i]$ if and only if (with $e_{i0}:=0$ and the abbreviation $L_i(T):=\ee^{r^\prime_0 T}F_i+ \sum_{j=1}^N \ee^{r_{ji} T}L_{ji} - \ee^{r_0 T}K_i$)
\begin{itemize}
	\item for all $i\in[N]$ with $n_i=1$ we have 
	\[ X_i(T)< L_i(T) - \sum_{j\in\pa_G(i)\cap\ndesc_G(i)} \ee^{r_{ij} T}L_{ij}\bone_{\{e_{jN_j}=0\}},  \]
	\item for all $i\in[N]$ with $n_i=N_i+1$ we have 
	\[ X_i(T) \geq L_i(T) - \sum_{j\in\pa_G(i)\cap\ndesc_G(i)} \ee^{r_{ij} T}L_{ij}\bone_{\{e_{jN_j}=0\}} - \sum_{j\in \pa_G(i)\cap \desc_G(i)} \ee^{r_{ij} T}L_{ij}\bone_{\{e_{j,N_i-1}=0\}},  \]
	\item and for all $i\in[N]$ with $2\leq n_i\leq N_i$ we have 
	\[ X_i(T)< L_i(T) - \sum_{j\in\pa_G(i)\cap\ndesc_G(i)} \ee^{r_{ij} T}L_{ij}\bone_{\{e_{jN_j}=0\}} - \sum_{j\in \pa_G(i)\cap\desc_G(i)} \ee^{r_{ij} T}L_{ij}\bone_{\{e_{j,n_i-1}=0\}}  \]
	and
	\[ X_i(T)\geq L_i(T) - \sum_{j\in\pa_G(i)\cap\ndesc_G(i)} \ee^{r_{ij} T}L_{ij}\bone_{\{e_{jN_j}=0\}} - \sum_{j\in \pa_G(i)\cap\desc_G(i)} \ee^{r_{ij} T}L_{ij}\bone_{\{e_{j,n_i-2}=0\}}.  \]
\end{itemize}
Thus, using the independence of $X_1(T),\ldots,X_N(T)$ and their log-normal distribution, we obtain on the one hand
\begin{align} &\bbp[D_{in}=e_{in}~\forall i\in[N],~n\in[N_i]]\nonumber\\
&\quad\quad= \prod_{i\colon n_i=1} \Phi(i,\bar\Si^\mathrm{m}_{in_i},T) \prod_{i\colon 2\leq n_i\leq N_i} (\Phi(i,\bar\Si^\mathrm{m}_{in_i},T)-\Phi(i,\bar\Si^\mathrm{m}_{i,n_i-1},T))\nonumber\\
&\quad\quad\quad\times \prod_{i\colon n_i=N_i+1} (1-\Phi(i,\bar\Si^\mathrm{m}_{i,n_i-1},T)),\label{joint}\end{align}
where $\bar\Si^\mathrm{m}_{in}:=\{j\in\pa_G(i)\cap\ndesc_G(i)\colon e_{jN_j}=0\}\cup \{j\in\pa_G(i)\cap \desc_G(i)\colon e_{j,n-1}=0\}$.

On the other hand, 
if we assume for a moment that formula \eqref{condprob-sdr} is valid, we deduce that for all $i\in[N]$ with $n_i=1$ we have
\begin{align*} \prod_{n=1}^{N_i} \bbp[D_{in}=e_{in}\mid \pa_{\bar G}(D_{in})=\pa_{\bar G}(e_{in})] &= \bbp[D_{i1}=1\mid\pa_{\bar G}(D_{in})=\pa_{\bar G}(e_{in})]\\
&= \Phi(i,\bar\Si^\mathrm{m}_{i1},T), 
\end{align*}
that for all $i\in[N]$ with $n_i=N_i+1$ we have
\begin{align*} \prod_{n=1}^{N_i} \bbp[D_{in}=e_{in}\mid \pa_{\bar G}(D_{in})=\pa_{\bar G}(e_{in})] &= \prod_{n=1}^{N_i} \bbp[D_{in}=0\mid \pa_{\bar G}(D_{in})=\pa_{\bar G}(e_{in})]\\
&=(1- \Phi(i,\bar\Si^\mathrm{m}_{i1},T)) \prod_{n=2}^{N_i}\frac{1-\Phi(i,\bar\Si^\mathrm{m}_{in},T)}{1-\Phi(i,\bar\Si^\mathrm{m}_{i,n-1},T)}\\
&= 1-\Phi(i,\bar\Si^\mathrm{m}_{iN_i},T),
\end{align*}
and that for all other $i\in[N]$ with $2\leq n_i\leq N_i$ we have
\begin{align*} &\prod_{n=1}^{N_i} \bbp[D_{in}=e_{in}\mid \pa_{\bar G}(D_{in})=\pa_{\bar G}(e_{in})] \\
&\quad\quad=\bbp[D_{in_i}=1\mid \pa_{\bar G}(D_{in_i})=\pa_{\bar G}(e_{in})]\prod_{n=1}^{n_i-1} \bbp[D_{in}=0\mid \pa_{\bar G}(D_{in})=\pa_{\bar G}(e_{in})]\\
&\quad\quad= \frac{\Phi(i,\bar\Si^\mathrm{m}_{in_i},T)-\Phi(i,\bar\Si^\mathrm{m}_{i,n_i-1},T)}{1-\Phi(i,\bar\Si^\mathrm{m}_{i,n_i-1},T)}(1-\Phi(i,\bar\Si^\mathrm{m}_{i1},T)) \prod_{n=2}^{n_i-1}\frac{1-\Phi(i,\bar\Si^\mathrm{m}_{in},T)}{1-\Phi(i,\bar\Si^\mathrm{m}_{i,n-1},T)}\\
&\quad\quad= \Phi(i,\bar\Si^\mathrm{m}_{in_i},T)-\Phi(i,\bar\Si^\mathrm{m}_{i,n_i-1},T).
\end{align*}
Comparing with \eqref{joint}, we conclude that
\begin{align*}
\bbp[D_{in}=e_{in}~\forall i\in[N],~n\in[N_i]]=\prod_{i=1}^N\prod_{n=1}^{N_i} \bbp[D_{in}=e_{in}\mid \pa_{\bar G}(D_{in})=\pa_{\bar G}(e_{in})], 
\end{align*}
which proves that the random variables $\{D_{in}\colon i\in[N],n\in[N_i]\}$ form a Bayesian network on $\bar G$.

It remains to demonstrate the correctness of the conditional probabilities stated in \eqref{condprob-sdr}. We only carry out the proof for $N_i>1$ and $n>1$; the case $n=1$ is simpler and can be achieved by a straightforward modification of the following arguments. It is evident by the monotonicity statement in (1) that $e_{i,n-1}=1$ immediately implies
\[ \bbp[D_{in}=1\mid \pa_{\bar G}(D_{in})=\pa_{\bar G}(e_{in})]=1,\]
so only the case $e_{i,n-1}=e_{i,n-2} = \ldots = e_{i1} = 0$ needs to be considered further. Conditional on $D_{i,n-1}=0$, we can now represent the parents of $D_{in}$ as
\[ \pa_{\bar G}(D_{in})=F_n(X_j(T)\colon j\in\anc_G(i)\setminus\{i\})\]
with some function $F_n$ whose explicit form can be derived from \eqref{Sin2}. Thus, abbreviating
\[ M_{in}:=L_i(T)-\sum_{j\in\pa_G(i)\cap\ndesc_G(i)} \ee^{r_{ij} T} L_{ij}\bone_{\{e_{jN_j}=0\}} - \sum_{j\in \pa_G(i)\cap \desc_G(i)} \ee^{r_{ij} T}L_{ij}\bone_{\{e_{j,n-1}=0\}}, \]
we obtain 
\begin{align*}
&\bbp[D_{in}=1\mid \pa_{\bar G}(D_{in})=\pa_{\bar G}(e_{in})] \\
&\quad\quad= \bbp[X_i(T)<M_{in} \mid X_i(T)\geq M_{i,n-1},~F_n(X_j(T)\colon j\in\anc_G(i)\setminus\{i\})=\pa_{\bar G}(e_{in})]\\
&\quad\quad=\bbp[X_i(T)<M_{in} \mid X_i(T)\geq M_{i,n-1}]
\end{align*}
by the independence of $X_1(T),\ldots,X_N(T)$. This is exactly what formula \eqref{condprob-mdr} states.
\halmos \epr

\bpr[of Theorem~\ref{Bayes}] The proof is completely analogous to that of Theorem~\ref{Bayes2}. \halmos\epr

\bpr[of Proposition~\ref{dsepeq}] (1)\quad We first assume that $V_1$ and $V_2$ are d-separated in $G$ and pick a chain between some $(i_1,N_{i_1})$ and $(i_k,N_{i_k})$ with $i_1\in V_1$ and $i_k\in V_2$. Observing that $(i,n)\to_{\bar G} (i^\prime,n^\prime)$ implies $i=i^\prime$ or $i\to_G i^\prime$, it follows that the chain under consideration cannot be a path (otherwise also $i_1$ and $i_k$ would be connected through a path in $G$, which contradicts d-separation). It can neither be of the form
\[ (i_1,N_{i_1}) \leftarrow_{\bar G} \ldots \leftarrow_{\bar G} (i_j,n_j) \to_{\bar G} \ldots \to_{\bar G} (i_k,N_{i_k}) \]
as this would entail a chain in $G$ of the same form, which would again violate the d-separation assumption. As a result, the chain in $\bar G$ must contain a type IV structure and is therefore blocked. The other direction follows similarly, we omit the details.

(2)\quad 
Every chain $(i_1,n_1)\darrow_{\bar G}\ldots\darrow_{\bar G}(i_k, n_k)$ with $i_1\in V_1$ and $i_k\in V_2$ induces a chain $i_1\darrow_G \ldots\darrow_G i_k$ in $G$, where consecutive copies of a vertex are understood as merged to a single vertex. By assumption, this chain in $G$ must be blocked by $V_0$ and therefore contain at least one blocking structure of type I--IV. If a blocking structure of type I, II or III is present, say, with middle vertex $i_j$, then $i_j\in V_0$ and the corresponding sequence in the original chain in $\bar G$ must contain a structure of the same type with some middle vertex $(i_j, n)$ and some $n\in [N_{i_j}]$, hence blocking the chain in $\bar G$. Similarly, for the last remaining case, if the chain in $G$ has a blocking structure of type IV, say, again with middle node $i_j$, then $V_0$ neither contains $i_j$ nor any of its descendants. Since this immediately implies that $\{(i,n)\colon i\in V_0,~n\in[N_i]\}$ and $\{(i,n)\colon i=i_j \text{ or } i\in\desc_G(i_j),~n\in[N_i]\}$ are disjoint, and the original chain in $\bar G$ necessarily has a type IV structure with some middle vertex $(i_j,n)$ and some $n\in[N_{i_j}]$, it is blocked by $\{(i,n)\colon i\in V_0,~n\in[N_i]\}$. \halmos
\epr

\bpr[of Proposition~\ref{SysImpProp}] (1) is obviously true. In (2) the statement for the absolute systemic impact follows from the definition, while it follows from (4) for the relative systemic impact. (3) is a well-known result, see Proposition~4.2 and the following remark in \cite{Levin09}. Next, (4) is a consequence of an elementary inequality: If $a/b$ and $c/d$ are bounded by $M$, then also $(a+c)/(b+d)$. Therefore, the maximum in \eqref{RSI-var} is attained on $J$. Finally, for (5) we notice that if $I_2$ and $J$ are d-separated in $G$ given $I_1$, then $\{(i,N_i)\colon i\in I_2\}$ and $\{(j,N_j)\colon j\in J\}$ are d-separated in $\bar G$ given $\{(i,n)\colon i\in I_1,~ n\in[N_i]\}$. Since observing $S_{I_1}=0$ means $S_{in}=0$ for all $i\in I_1$ and $n\in[N_i]$, the claim follows from the fact that d-separation entails independence in Bayesian networks.\halmos\epr

\end{appendix}

\subsection*{Acknowledgements}
We are grateful to Rama Cont for enlightening discussions on systemic risk and to the Isaac Newton Institute of the University of Cambridge for its hospitality. 

\addcontentsline{toc}{section}{References}
\bibliographystyle{plainnat}
\bibliography{bib-DefaultBonds}

\end{document}